\documentclass[11pt,tightenlines,eqsecnum,floats,aps,amsmath,amssymb,nofootinbib,prd,shownopacs,floatfix]{revtex4}

\usepackage{graphicx}
\usepackage{epstopdf}
\usepackage{latexsym}
\usepackage{amssymb}
\usepackage{amsmath}
\usepackage{color}
\usepackage{mathrsfs}
\usepackage{xparse}
\usepackage{float}
\usepackage{mathtools}
\usepackage{multirow}
\usepackage[center]{subfigure}

\usepackage{xcolor}
\begin{document}
\renewcommand\arraystretch{2}
 \newcommand{\bq}{\begin{equation}}
 \newcommand{\eq}{\end{equation}}
 \newcommand{\bqn}{\begin{eqnarray}}
 \newcommand{\eqn}{\end{eqnarray}}
 \newcommand{\nb}{\nonumber}
 \newcommand{\cb}{\color{blue}}
    \newcommand{\cc}{\color{cyan}}
     \newcommand{\lb}{\label}
        \newcommand{\cm}{\color{magenta}}
\newcommand{\rc}{\rho^{\scriptscriptstyle{\mathrm{I}}}_c}
\newcommand{\rd}{\rho^{\scriptscriptstyle{\mathrm{II}}}_c}
\NewDocumentCommand{\evalat}{sO{\big}mm}{%
  \IfBooleanTF{#1}
   {\mleft. #3 \mright|_{#4}}
   {#3#2|_{#4}}%
}
\newcommand{\PRL}{Phys. Rev. Lett.}
\newcommand{\PL}{Phys. Lett.}
\newcommand{\PR}{Phys. Rev.}
\newcommand{\CQG}{Class. Quantum Grav.}
\newcommand{\parallelsum}{\mathbin{\!/\mkern-5mu/\!}}

\title{Love Numbers of Covariant Loop Quantum Black Holes}
\author{Meysam Motaharfar$^1$}
\email{mmotah4@lsu.edu}
 \author{Parampreet Singh$^{1,2}$}
\email{psingh@lsu.edu}
\affiliation{$^1$ Department of Physics and Astronomy, 
Louisiana State University, Baton Rouge, LA 70803, USA}
 \affiliation{$^2$ Center for Computation and Technology,
Louisiana State University, Baton Rouge, LA 70803, USA}

\begin{abstract}
We investigate the linear static response of three covariant loop quantum black holes, namely, the two models proposed by Zhang, Lewandowski, Ma, and Yang (ZLMY) and the Alonso-Bardaji, Brizuela, and Vera (ABV) model, to an external tidal field. Using effective spacetime description, we uniquely extract the tidal Love numbers (TLNs) using perturbative solutions derived from the Green's function technique. Our findings reveal that, in contrast to the classical Schwarzschild black hole, the TLNs for loop quantum black holes are generally nonzero. The sign and magnitude of the TLNs depend on the spin of the external tidal field, the multipole number, and the details of the loop quantized model. The magnitude of the TLNs is found to be Planck scale suppressed for all three models, implying a stronger tidal deformability for black holes with Planckian mass. Additionally, for the same black hole mass, the magnitude of the TLNs for the ABV model is larger than the ZLMY models. We also find that the TLNs exhibit logarithmic running behavior at the leading order, even for low multipole numbers, in response to scalar and vector field perturbations. These distinct features of the TLNs can serve as a potential tool to differentiate between various quantization ambiguities arising in loop quantum black holes. We briefly discuss the potential phenomenological and theoretical implications of nonzero TLNs for black hole physics.
\end{abstract}

\maketitle

\section{Introduction} 

One of the most fascinating predictions of Einstein’s general relativity (GR) is the emission of gravitational waves from binary systems. The direct detection of gravitational waves by the LIGO/Virgo interferometers \cite{LIGOScientific:2016aoc}, along with the future observations by the planned Laser Interferometer Space Antenna (LISA) from the coalescence of binary systems, such as binary black holes, binary neutron stars, and binary neutron-black hole systems, opens unprecedented opportunities to reveal the nature of gravity in the strong-field regime. During the inspiral phase of binary systems, the orbital separation of the binary shrinks because of the emission of gravitational waves, and tidal interactions become relevant. Among these tidal interactions, the tidal deformability of an object acted upon by the external gravitational field of its companion plays an important role in the dynamics of the binary system. This deformability influences the waveform of the emitted gravitational waves at the 5th post-Newtonian order \cite{Flanagan:2007ix, Vines:2010ca, Narikawa:2023deu} and encodes information on the internal structure of the compact object and the nature of gravity and matter. For instance, tidal deformability offers critical insights into the equation of state of neutron stars \cite{Flanagan:2007ix, Baiotti:2016qnr, LIGOScientific:2018cki} and enables probing the presence of new physics near the event horizon of black holes \cite{Maselli:2018fay, Datta:2021hvm}.

Tidal deformability of a self-gravitating object under the influence of an external tidal field is characterized by a set of complex coefficients that capture the conservative and dissipative aspects of the response. While the conservative part, commonly referred to as tidal Love numbers (TLNs) \cite{10.1093/mnras/69.6.476}, encodes information about how the object would deform in response to the external tidal field, the dissipative part describes the amount of energy that would be lost due to tidal interactions. With vanishing TLNs in four-dimensional GR, black holes are the most resilient compact objects in the cosmos. Explicit calculations in general relativity have shown that several families of asymptotically flat black holes in four-dimensional spacetime have exactly vanishing TLNs. This result was first established for Schwarzschild black holes \cite{Schwarzschild, Gurlebeck:2015xpa, Kol:2011vg, Hui:2020xxx, Bhatt:2023zsy}, and later it was extended to Kerr \cite{LeTiec:2020spy, Kerr}, Reissner-N\"ordstrom \cite{Cardoso:2017cfl, Pereniguez:2021xcj, Rai:2024lho}, and Kerr-Newman \cite{Ma:2024few} black holes. Recent studies have further confirmed that static TLNs also vanish, even when taking into account the effect of non-linearities in the external tidal field perturbations \cite{non-linear TLNs, Combaluzier-Szteinsznaider:2024sgb, Gounis:2024hcm}. Interestingly, this result holds only in four dimensions, hinting at the presence of a hidden ladder symmetry for spherically symmetric spacetimes in four-dimensional GR \cite{symmetry of TLNs, Combaluzier-Szteinsznaider:2024sgb, Gounis:2024hcm}. In contrast, higher-dimensional black holes \cite{Hui:2020xxx, Kol:2011vg, Ma:2024few, Pereniguez:2021xcj, higher dimension TLNs} in general display nonzero TLNs. Similarly, nonzero TLNs also emerge in the presence of a cosmological constant \cite{Nair:2024mya, Franzin:2024cah}, in extended gravitational theories \cite{Cardoso:2017cfl, Cardoso:2018ptl, DeLuca:2022tkm, Barura:2024uog, Katagiri:2024fpn, Diedrichs:2025vhv}, and when matter is in the environment surrounding the objects \cite{Cardoso:2019upw, DeLuca:2021ite, DeLuca:2022xlz, Katagiri:2023yzm, Cannizzaro:2024fpz}. The detection of nonzero TLNs serves as evidence of new physics, offering an opportunity to constrain quantum corrections of the event horizon and providing a powerful tool for testing theories of gravity in the strong-field regime.

While black holes are among the most compelling confirmations of GR within its domain of validity, the central singularity of black holes, along with the big bang singularity, indicate that classical GR breaks once spacetime curvature approaches the Planck regime. It has long been expected that an ultimate theory of quantum gravity would resolve classical singularities, thus enabling us to make physical predictions at the Planck scale. In the last decade, techniques from loop quantum gravity, a candidate theory of quantum gravity, have been successfully applied to understand resolution of singularities in a wide range of settings, including cosmological \cite{Ashtekar:2011ni} and black hole \cite{Ashtekar:2023cod} spacetimes.  The key prediction of the loop quantization of all the spacetimes studied so far is that quantum geometric effects resolve the classical singularities, such as replacing the big bang with a big bounce \cite{Ashtekar:2006rx,Ashtekar:2006wn,Ashtekar:2007em}. Although the spacetime geometry is fundamentally discrete, governed by a quantum difference equation, it turns out that the underlying quantum dynamics are accurately captured by an effective continuum spacetime description \cite{Diener:2014mia,Diener:2017lde,Singh:2018rwa} commonly derived by a polymerization technique. Assuming the validity of effective dynamics, the generic resolution of strong curvature singularities in both isotropic and anisotropic spacetimes, including the Kantowski-Sachs spacetime, has been extensively investigated in the literature \cite{singularity-resolution}. Therefore, the primary focus in loop quantum black hole studies is on effective models where the effective Hamiltonian constraint is directly constructed from polymerization of the classical Hamiltonian. These models include those homogeneous minisuperspace models benefiting from the isometry between the interior of the Schwarzschild black hole and Kantowski-Sachs spacetime \cite{Ashtekar:2005qt, Modesto:2005zm, Chiou:2008nm, Dadhich:2015ora, Corichi:2015xia, Olmedo:2017lvt, Ashtekar:2018lag, Ashtekar:2018cay, Bodendorfer:2019nvy, Zhang:2020qxw, ElizagaNavascues:2022rof, Ongole:2023pbs, Ongole:2025lti}. Similar studies have also been carried out in the midisuperspace setting \cite{LQBH, Alonso-Bardaji:2021yls, Zhang:2024khj, Zhang:2024ney} (see also Ref. \cite{Rovelli:2024sjl} for an overview of other approaches inspired from loop quantizations).

The static response of loop quantum black holes to an external tidal field, was studied for the first time by the authors in Ref. \cite{Motaharfar:2025typ} for the case of Ashtekar-Olmedo-Singh (AOS) model \cite{Ashtekar:2018lag, Ashtekar:2018cay}. It was found that the TLNs are, in general, nonzero and negative, in contrast to classical theory, which is vanishing, for all three scalar, vector, and axial gravitational responses and all multipole numbers. Moreover, the magnitude of the TLNs is Planck scale suppressed, implying that the TLNs for astrophysical black holes are small, while they are stronger for Planckian mass black holes. Such Planck scale suppression, albeit with different scaling, was also found in Ref. \cite{Kim:2020dif}, where quantum corrections to the TLNs for Schwarzschild black holes are obtained from a sum rule for the TLNs derived from quantum field theory computations in a semiclassical regime. We also found that the AOS model with $M\gtrsim 4.3 \times 10^{4}M_{\textrm{Pl}}$ is consistent with current and next-generation detection limits for TLNs \cite{Pani:2019cyc}. Since the AOS model is primarily designed for the quantization of macroscopic black holes, our results demonstrate that it has a much wider range of validity. Our results also suggest that the quantum deformability of loop quantum black holes due to the inherent fuzziness of the underlying quantum geometry reveals a fundamentally distinct internal structure compared to their classical counterparts. In other words, loop quantum black holes bypass the realistic version of the no-hair theorem \cite{Gurlebeck:2015xpa} and obtain a quantum hair accessible to outside observers via the TLNs, leading to profound theoretical and phenomenological implications.

An important question for loop quantum, symmetry-reduced black hole models is preserving general covariance while transitioning from the classical theory to the effective quantum theory. While general covariance is guaranteed by the diffeomorphism invariance of the action in the Lagrangian formulation of GR, establishing general covariance in the Hamiltonian formulation is more challenging due to the $3+1$ decomposition of spacetime. To ensure general covariance in loop quantum black holes, the spacetime diffeomorphism gauge is typically fixed by choosing a specific matter field, and then the Hamiltonian is quantized by applying LQG techniques. However, the resulting effective Hamiltonian constraint upon quantization is valid only within the preferred gauge. This indicates that by deviating from the preferred gauge, the effective Hamiltonian constraint needs to be modified, ensuring the resulting effective metric is diffeomorphically equivalent to the one obtained in the preferred gauge. To address this issue, various effective models have been constructed by applying loop quantization techniques to spherically symmetric spacetime in midisuperspace to preserve general covariance \cite{LQBH, Alonso-Bardaji:2021yls, Zhang:2024khj, Zhang:2024ney}. Therefore, it is interesting to explore the linear static response of covariant loop quantum black holes to an external tidal field, namely a scalar field, a vector field, and an axial gravitational field, and extract quantum gravitational corrections to the TLNs.

To explore the impact of quantum gravitational effects on TLNs, we focus on covariant loop quantum black hole models proposed by Zhang, Lewandowski, Ma, and Yang (ZLMY) \cite{Zhang:2024khj, Zhang:2024ney} as well as the one developed by Alonso-Bardaji, Brizuela, and Vera (ABV) \cite{Alonso-Bardaji:2021yls}, which are constructed to maintain general covariance independent of the gauge choices. In the former, the authors proposed a method to relax gauge fixing condition and derived an effective Hamiltonian constraint that remains valid uniformly across different gauges. More precisely, for spherically symmetric spacetime, by introducing a freely chosen function in the construction of the effective metric and imposing minimal requirements on an arbitrary effective Hamiltonian constraint,  an effective Hamiltonian constraint that preserves general covariance is derived. Then two different effective masses that satisfy the covariance equations are proposed from which one derives the corresponding effective metrics. In both solutions, the classical singularity within the black hole interior is replaced by a smooth transition surface. This surface connects a trapped region in the past to an anti-trapped region in the future, effectively extending the Schwarzschild interior to include a structure interpreted as a white hole horizon. One of the solutions exhibits a double-horizon structure similar to the Reissner-Nordstr\" om black hole with a small charge, which will hereafter be referred to as the ZLMY-I model, and the other solution is referred to as the ZLMY-II model. Although these two models share certain similarities, their quasi-normal spectra \cite{Konoplya:2024lch}, shadow \cite{Liu:2024soc}, and gravitational lensing effects \cite{Liu:2024wal, Wang:2024iwt} are distinct. Moreover, using the Newman-Janis algorithm, rotating versions of these models have been constructed, and the impact of quantum gravitational effects on their shadow has been studied in Ref. \cite{Ban:2024qsa}. On the other hand, in the ABV model, the authors include the anomaly-free holonomy corrections through a canonical transformation and a linear combination of constraints of GR, ensuring the closure of the constrained algebra independent of the gauge choice on the phase space. As opposed to the standard polymerization procedure, where only the extrinsic curvature is replaced with a trigonometric function, in this case both the extrinsic curvature and its conjugate are polymerized. The resulting effective spacetime has a singularity-free interior, where the singularity is replaced with a transition surface joining two asymptotically flat exterior regions of equal mass, similar to other loop quantum black holes. Its quasi-normal spectra have been studied in Refs. \cite{Moreira:2023cxy, Gingrich:2024tuf, Bolokhov:2023bwm}, while some peculiar properties in quasi-normal modes for higher overtones are reported in Ref. \cite{Fu:2023drp}, and the gravitational lensing effects are studied in Refs. \cite{Junior:2023xgl, Soares:2023uup}. Given these models, it is relevant to explore the static response of the aforementioned covariant loop quantum black holes, and to answer the following questions: i) Do loop quantum black holes deform in general? ii) Could loop quantum black holes leave detectable observational effects while they are internally consistent? Or, put differently, how universal is the Planck scale suppression of the TLNs? iii) Could we use the TLNs to differentiate quantum ambiguities arising in loop quantum black holes?

To address these questions, we study the linear static response of covariant loop quantum black holes, the ZLMY-I model, the ZLMY-II model, and the ABV model, to an external tidal field, namely, a scalar field, a vector field, and an axial gravitational field. Assuming that quantum parameters are small and the validity of effective metric, we uniquely extract the TLNs for the three lowest multipole numbers $\ell$ from perturbative solutions derived via the Green's function technique up to the third order in the perturbation series expansion. In general, we find that the TLNs are nonzero, and the magnitude and sign of the TLNs depend on the spin of the external tidal field, the multipole number, and the specified loop quantum black hole model. Specifically, we find that the magnitude of the TLNs is Planck scale suppressed for all three models, indicating a stronger tidal deformability for black holes with a Planckian mass. Moreover, for the same black hole mass, the ABV model has larger TLNs in comparison with the ZLMY-I/II models. For the leading order quantum gravitational corrections to the TLNs, our results can be briefly summarized as follows. For scalar field responses, we find that the TLNs are negative for the three lowest multipole numbers and for all three black hole models except $\ell=0$ for the ZLMY-I/II models, which have vanishing TLNs in this case. However, it is found that these models predict positive, negative, or zero logarithmic running for different multipole numbers, making them distinct. For the vector field responses, our results show that the TLN is negative for $\ell=1$, while it is positive for the other two next multipole numbers, $\ell=2$ and $\ell=3$, in the case of the ZLMY-I model. In the case of the ZLMY-II model, the first two lowest multipole numbers exhibit a negative sign for the TLN, while the TLN has a positive sign for the next multipole number. On the other hand, the vector TLNs are positive for all three lowest multipole numbers in the case of the ABV model. Although the ZLMY-I/II models predict positive running for the vector TLNs, the vector TLNs for the ABV model exhibit no logarithmic running. Finally, our results show that the tensor TLNs are all positive with positive running for the three lowest multipole numbers and for all the three models. They also do not exhibit logarithmic running.

The outline of the manuscript is as follows. In Section \ref{Section II}, we briefly review three covariant loop quantum black holes, namely, the ZLMY-I/II and ABV models, and present the effective metric describing the underlying quantum dynamics. In Section \ref{Section III}, we revisit the definition of TLNs and then discuss the Green's function technique to derive perturbative solutions and briefly review the extraction procedure for TLNs. In Section \ref{Section IV}, we compute the TLNs in response to a scalar field, a vector field, and an axial gravitational field perturbation for the three covariant loop quantum black holes discussed in Section \ref{Section II}. Finally, we summarize the results and present our conclusions in Section \ref{Section V}. We use Planck units for the discussion of results.

\section{Brief review of covariant loop quantum black holes} \label{Section II}

In this section, we briefly review three covariant loop quantum black holes proposed in Refs. \cite{Zhang:2024khj, Zhang:2024ney, Alonso-Bardaji:2021yls}, which preserve the general covariance independent of gauge choice. We then present the effective metric describing the underlying quantum dynamics of these models and briefly discuss their properties.

\subsection{Zhang, Lewandowski, Ma, and Yang models}

In Refs. \cite{Zhang:2024khj, Zhang:2024ney}, Zhang, Lewandowski, Ma, and Yang (ZLMY) proposed a method to relax gauge fixing conditions and obtained an effective Hamiltonian constraint that preserves general covariance and remains valid uniformly across different gauges. For spherically symmetric spacetimes, by introducing a freely chosen function in the construction of the effective metric and imposing minimal requirements on an arbitrary effective Hamiltonian constraint, an effective Hamiltonian constraint that preserves general covariance was proposed as:
\begin{align}
    H_{\textrm{eff}} = -2 E^{\phi} \left(\partial_{s_{1}}M_{\textrm{eff}} + \frac{\partial_{s_{2}}M_{\textrm{eff}}}{2} s_{3} + \frac{\partial_{s_{4}} M_{\textrm{eff}}}{s_{4}}s_{5} + \mathcal{R}\right),
\end{align}
where $\mathcal{R}(s_{1}, M_{\textrm{eff}})$ is an arbitrary function, and the effective mass $M_{\textrm{eff}}(s_{1}, s_{2}, s_{4})$ and $\mu$ satisfy the following equations
\begin{align}
    \frac{\mu s_{1}s_{4}}{4} &= \left(\partial_{s_2} M_{\textrm{eff}} \right)\partial_{s_2} \partial_{s_{4}} M_{\textrm{eff}} - (\partial_{s_{4}}M_{\textrm{eff}}) \partial^{2}_{s_{2}}M_{\textrm{eff}}, \label{Meff-1}\\ &
    (\partial_{s_{2}}\mu)\partial_{s_{4}}M_{\textrm{eff}}-(\partial_{s_{2}}M_{\textrm{eff}})\partial_{s_{4}}\mu = 0\label{Meff-2} .
\end{align}
Here $\mu$ is an arbitrary function introduced in the effective metric and $s_{i}$ denotes scalar quantities given by
\begin{align}
    s_{1} &= E^{x},  \ \ \ \ s_{2} = K_{\phi},  \ \ \ \ \ s_{3} = \frac{K_{x}}{E^{\phi}}, \\ s_{4} & = \frac{\partial_{x} E^{x}}{E^{\phi}},  \ \ \ \ s_{5} = \frac{1}{E^{\phi}} \partial_{x}\left(\frac{\partial_{x}E^{x}}{E^{\phi}}\right),
\end{align}
where $E^{x}$ and $E^{\phi}$ are triads and $K_{x}$ and $K_{\phi}$ are their conjugate momentum, i.e., extrinsic curvatures, respectively. Two solutions for loop quantum black holes in Ref. \cite{Zhang:2024khj} were found by applying the polymerization techniques, by replacing the extrinsic curvature with its trigonometric function. It was found that the following effective mass satisfies Eq. (\ref{Meff-1}), ensuring general covariance of the effective metric
\begin{align}\label{Meff1}
    M_{\textrm{eff}}^{(1)} = \frac{\sqrt{s_{1}}}{2} +  \frac{s_{1}\sqrt{s_{1}}\sin^2(\frac{\xi s_{2}}{\sqrt{s_{1}}})}{2\xi^2} - \frac{\sqrt{s_{1}}(s_{4})^2}{8}e^{\frac{{2i\xi s_{2}}}{\sqrt{s_{1}}}},
\end{align}
and it is a solution to Eq. (\ref{Meff-2}) for $\mu = \mu_{1}=1$. Notably, even though effective mass (\ref{Meff1}) is complex, resulting in a complex effective Hamiltonian constraint, the effective metric turns out to be real if we allow $K_{x}$ and $K_{\phi}$ to take complex values as noted in Ref. \cite{Zhang:2024khj}. Given the effective Hamiltonian constraint, one can find the effective metric in a familiar static, spherically symmetric form
\begin{align}\label{spherically-symmetric-metric}
    \mathrm{d} s^2_{1} = - A_{1}(r)\mathrm{d}t^2 + B_{1}(r)\mathrm{d}r^2 + r^2 \mathrm{d} \Omega^2,
\end{align}
with $d\Omega^2 = d\theta^2 + \sin^2(\theta) d\phi^2$. The metric functions read as
\bq\label{model-I}
A_{1}(r) = B_{1}(r) = f(r) \left[1+  \frac{\xi^2}{r^2}f(r)\right], \ \ \ \ \ \ \ \ \ \ \ f(r) = 1-\frac{r_{H}}{r},
\eq
where $r_{H}=2M$ with $M$ being the black hole mass, and $\xi$ is a regularization parameter of quantum origin. This parameter is fixed by quantum geometry as $\xi = \sqrt{\Delta}$ with $\Delta = 4\sqrt{3}\pi \gamma l_{\textrm{Pl}}^2$ being the minimum eigenvalue of the area operator in LQG. Here $l_{\textrm{Pl}}$ and $\gamma\simeq 0.2375$ are, respectively, the Planck length and the Barbero-Immirzi parameter fixed by black hole thermodynamics in LQG. The metric function $A_{1}(r)$ has two real positive roots for all $M>0$ given by $r_{+} = r_{H}$ and $r_{-} = {\xi^2}/{\beta}-{\beta}/{3}$, with $\beta^3 = 3 \xi^2(\sqrt{81M^2 + 3 \xi^2}-9M)$. This implies that the effective metric (\ref{model-I}) exhibits double-horizon structure, namely, an event horizon $r_{+}$ and Cauchy horizon $r_{-}$, similar to Reissner-Nordstr\" om with a small charge. Such casual structure has been observed in loop quantum black holes where the classical singularity at the interior of a black hole is replaced with a smooth transition surface joining two asymptotically flat regions \cite{Ashtekar:2018lag, Ashtekar:2018cay}. Hereafter, we will refer to this model as the ZLMY-I model.

The second effective mass is given by 
\begin{align}
    M_{\textrm{eff}}^{(2)} = \frac{\sqrt{s_{1}}}{2} +  \frac{s_{1}\sqrt{s_{1}}\sin^2(\frac{\xi s_{2}}{\sqrt{s_{1}}})}{2\xi^2} - \frac{\sqrt{s_{1}}(s_{4})^2 \cos^2(\frac{\xi s_{2}}{\sqrt{s_{1}}})}{8},
\end{align}
which is the solution of Eq. (\ref{Meff-2}) for $\mu = \mu_{2} = 1 + \xi^2 (\sqrt{s_{1}} - 2 M_{\textrm{eff}}^{(2)})/({s_{1}\sqrt{s_{1}}})$ and the metric functions of the effective metric read as
\begin{align}\label{model-II}
A_{2}(r) = f(r), \ \ \ \ \ \ \ \ \ \  
B_{2}(r)  =  f(r) \left[1+  \frac{\xi^2}{r^2}f(r)\right],
\end{align}
where again the classical singularity is replaced by a smooth transition surface connecting the black hole region to the white hole region. Hereafter, we will refer to this model as the ZLMY-II model. As it is obvious from the effective metrics (\ref{model-I}) and (\ref{model-II}), they quantify the deviation from the classical Schwarzschild black hole spacetime through the quantum parameter $\xi$, while recovering the classical regime in the limit $\xi\rightarrow 0$. 

\subsection{Alonso-Bardaji, Brizuela, and Vera model}

In Ref. \cite{Alonso-Bardaji:2021yls}, Alonso-Bardaji, Brizuela, and Vera (ABV) proposed a covariant effective model ensuring general covariance by requiring the closure of deformed constrained algebra using following canonical transformations:
\begin{align}\label{ABV-transfromation}
    E^{x} \rightarrow \tilde{E}^{x}, \ \ \ \ \ \ \ \ \ \ \ K_{x} \rightarrow \tilde{K}_{x}, \ \ \ \ \ \ \ \ \ \ \
 E^{\phi} \rightarrow \frac{\tilde{E}^{\phi}}{\cos(\lambda \tilde{K}_{\phi})},  \ \ \ \ \ \ \ \ \ \ \ K_{\phi} \rightarrow \frac{\sin (\lambda \tilde{K}_{{\phi}})}{\lambda} .
\end{align}
Here $\lambda$ is the polymerization parameter of quantum origin and for some real $\lambda\neq 0$ leaves the $(K_{x}, E^{x})$ pair invariant. This transformation introduces a trigonometric function of $K_{\phi}$ to include holonomy corrections through polymerization. Note that, as opposed to the usual polymerization procedure, not only the variable $K_{\phi}$ but also its conjugate $E^{\phi}$ is transformed, guaranteeing that the deformed constrained algebra remains the same. The parameter $\lambda$ is a positive dimensionless constant, and in the limit $\lambda\rightarrow0$, classical GR is recovered. This transformation leaves the diffeomorphism constraint invariant. As long as $\cos(\lambda \tilde{K}_{\phi})$ does not vanish, the dynamical content of the theory is same as GR. Since the Hamiltonian constraint diverges when $\cos(\lambda \tilde{K}_{\phi})$ vanishes, it is necessary to perform a linear combination between the Hamiltonian constraint and the diffeomorphism constraint to ensure an algebra free of anomalies \cite{Alonso-Bardaji:2021yls}. Taking all of these into consideration, the deformed Hamiltonian constraint is found to be
\begin{align}
    \nonumber \tilde{\mathcal{H}} & = -\frac{\tilde{E}^{\phi}}{2\sqrt{\tilde{E}^{x}}\sqrt{1+\lambda^2}}\left(1 + \frac{\sin^2(\lambda \tilde{K}_{\phi})}{\lambda^2}\right) - \sqrt{\tilde{E}^x}\tilde{K}_{x} \frac{\sin(2\lambda \tilde{K}_{\phi})}{\lambda \sqrt{1+\lambda^2}}\left(1+\left(\frac{\lambda {\tilde{E}^{x\prime}}}{2 \tilde{E}^{\phi}}\right)^2\right) \\ & + \frac{\cos^2(\lambda \tilde{K}_{\phi})}{2\sqrt{1+\lambda^2}}\left(\frac{\tilde{E}^{x\prime}}{2E^{\phi}} \left(\sqrt{\tilde{E}^{x}}\right)^{\prime} + \sqrt{\tilde{E}^{x}} \left(\frac{{\tilde{E}^{x\prime}}}{\tilde{E}^{\phi}}\right)^{\prime}\right) .
\end{align}
Here the prime denotes the derivative with respect to $x$. The deformed Hamiltonian constraint recovers GR in the limit $\lambda\rightarrow 0$ since the canonical transformation (\ref{ABV-transfromation}) is identical. For the deformed Hamiltonian constraint, one can find that the following expression
\begin{align}
    M = \frac{\sqrt{\tilde{E}^{x}}}{2}\left(1+\frac{\sin^2(\lambda \tilde{K}_{\phi})}{\lambda^2} - \left(\frac{\tilde{E}^{x\prime}}{2\tilde{E}^{\phi}}\right)^2\cos^2(\lambda \tilde{K}_{\phi})\right),
\end{align}
commutes with the deformed Hamiltonian constraint,  is a constant of motion, and it can be interpreted as the mass of a black hole as discussed in Ref. \cite{Alonso-Bardaji:2021yls}. By solving the equations of motion, which are derived from the Poisson brackets of phase space variables with the Hamiltonian constraint, one can obtain an explicitly spherically symmetric metric with the following metric functions
\begin{align}\label{model-III}
    A_{3}(r) = f(r ) = 1- \frac{r_{H}}{{r}}, \ \ \ \  \ \ \ \ \ B_{3}(r) = \left(1- \frac{r_{0}}{r}\right)f(r),
\end{align}
where $r_{0} = {2 M \lambda^2}/(1+\lambda^2) $. For $\lambda \neq 0$ and assuming $M>0$, it follows that $0<r_{0}<2M$. While, for $\lambda=0$, $r_{0}$ goes to zero, and the metric for the classical Schwarzschild black hole is recovered. It important to note that $\lambda$ is constant over any dynamical trajectory but it varies for each different solution as the black hole mass changes. The model in this sense has some similarity with AOS model. From a physical perspective, it seems more natural to fix the area gap to be $\Delta = 4\pi r_{0}^2$, i.e., $r_{0}/(2M) = \lambda^2/(1+\lambda^2) = \sqrt{\Delta/(16\pi)}M^{-1}$, from which one obviously infers that $\lambda$ is mass dependent \cite{Alonso-Bardaji:2024tvp}. The effective spacetime (\ref{model-III}) is free from central singularity due to holonomy corrections which are captured by the polymerization of phase space variables. The singularity is replaced by a smooth spacelike transition surface. This transition surface joins two exterior, asymptotically flat regions with equal mass, similar to the case in the ZLMY-I/II models. Hereafter, this model is referred to as the ABV model in this study.

It should be noted that the effective metrics (\ref{model-I}), (\ref{model-II}), and (\ref{model-III}), have the same relation between the event horizon and the black hole mass, i.e., $r_{H}=2M$, as in the classical Schwarzschild black hole. As we will discuss in Sections \ref{Section III} and \ref{Section IV}, this feature enables us to transform the equation for the radial component of perturbations in such a way that all the quantum gravitational effects are captured in the effective potential. Then, we can compute quantum gravitational corrections to the classical TLNs for the ZLMY-I/II and ABV black hole models by finding perturbative solutions via the Green's function technique, while assuming that quantum parameters are small for macroscopic black holes. Therefore, we will briefly review the definition of TLNs, and the Green's function technique to extract TLNs in the next section. 

\section{Tidal Love numbers:  Green's Function Technique} \label{Section III}

In this section, we briefly revisit the definition of TLNs in GR and review the procedure to extract TLNs for a spherically symmetric spacetime by applying the Green's function technique developed in Ref. \cite{Barura:2024uog}. Assuming that the external tidal field is weak, the linear perturbation theory is valid for studying the tidal response of a self-gravitating object. Due to the spherical symmetry of the spacetime, the angular dependence of field perturbation can be separated from time and radial dependence by decomposing field perturbations in terms of spherical harmonics. For a general spherically symmetric metric given by
\begin{align}\label{sg}
     \mathrm{d} s^2 = - A(r)\mathrm{d}t^2 + B(r)\mathrm{d}r^2 + r^2 \mathrm{d} \Omega^2,
\end{align}
the radial component of spin $s$ field perturbation, $\Psi(r)$, namely a scalar ($s=0$), a vector ($s=1$), and an axial gravitational field perturbation ($s=2$), is governed by  a second order equation. At the linear order in field perturbation, this equation is given by
\bq\label{perturbation-r}
\sqrt{A(r)B(r)} \frac{\mathrm{d}}{\mathrm{d}r}\left(\sqrt{A(r)B(r)}\frac{\mathrm{d}}{\mathrm{d}r}\Psi(r)\right) + \left[\omega^2- A(r)V(r)\right] \Psi(r) = 0,
\eq
where $\omega$ is the frequency of the external tidal field. The effective potential  reads as \cite{Arbey:2021jif, Konoplya:2024lch}
\begin{equation}\label{effective potential}
V(r) = 
\frac{\ell (\ell + 1)}{r^2} 
+ \frac{1 - s}{2 rA(r)} \frac{d[A(r)B(r)]}{dr}
+ \frac{s(s - 1)[B(r) - 1]}{r^2},
\end{equation}
where $\ell$ and $s$ are the multipole number and spin of the external tidal field, respectively, and $\ell\ge s$. Note that we have factored out the function $A(r)$ from the effective potential, $V(r)$, following the notation in Ref. \cite{Barura:2024uog}. We should also point out that, under the Regge-Wheeler gauge, axial gravitational field perturbations in four-dimensional GR result in the same effective potential as described in Eq. (\ref{effective potential}) with $s=2$ \cite{Katagiri:2023umb}. Hereafter, we use the effective potential (\ref{effective potential}) with $s=2$ to study the axial gravitational field perturbation.

The tidal response of a self-gravitating object under the influence of an external tidal field is characterized by a set of complex numbers, which contains the conserved and dissipative parts. The conserved part (which is also the real part), the so-called TLNs, describes the way the object would deform. On the other hand, the dissipative part (imaginary part) quantifies the amount of energy that would be dissipated due to tidal interactions. More precisely, TLNs measure the induced moments acquired by the object in  presence of a static external tidal field. Assuming that the amplitude of the field is small and its wavelength is much larger than the horizon scale, the field perturbation in black hole spacetime geometry is then described by a static linear perturbation theory (zero-frequency limit), i.e., $\omega=0$. Hence, for the classical Schwarzschild black hole, $A(r)=B(r) = f(r) = 1 - r_{H}/r$, and from Eq. (\ref{perturbation-r}) in the static limit, we have
\bq\label{perturbations-GR}
f(\tilde{r}) \frac{\mathrm{d}}{\mathrm{d}\tilde{r}}\left(f(\tilde{r})\frac{\mathrm{d}}{\mathrm{d}\tilde{r}}\Psi(\tilde{r})\right)  - f(\tilde{r}) V_{\textrm{GR}}(\tilde{r}) \Psi(\tilde{r}) = 0.
\eq
Here we have defined $\tilde{r} = r_{H}/r$ for convenience.
The effective potential $V_{\textrm{GR}}$ for the Schwarzschild black hole in four-dimensional GR also reads as
\begin{align}\label{GR-effective-potential}
    V_{\textrm{GR}}(\tilde{r}) = \frac{\ell(\ell+1)}{\tilde{r}^2} + \frac{(1-s^2)}{\tilde{r}^3},
\end{align}
In terms of $\tilde r$, the horizon is now located at $\tilde{r}=1$. The general solution to Eq. (\ref{perturbations-GR}) can be written in terms of a hypergeometric function, and it is straightforward to find a solution that is regular at the horizon. For asymptotically flat metric, i.e., $f(r)\rightarrow 1$, the large-$r$ behavior of the solution is in general
given by \cite{Katagiri:2023umb}
\bq \label{TLNs}
\Psi(r) \sim  \tilde{r}^{\ell+1} \left[1 + \mathcal{O}\left(\tilde{r}^{-1}\right)\right] + \kappa_{\ell}^{s}  \tilde{r}^{-\ell} \left[1 + \mathcal{O}\left(\tilde{r}^{-1}\right)\right],
\eq
up to a constant $\kappa_{\ell}^{s}$, which is the so-called TLNs quantifying the tidal deformability of the object. The solution contains a growing mode, which can be regarded as the external tidal field, and a decaying mode, which is understood as the induced response. It turns out that within GR, the solution for the Schwarzschild metric only has a growing mode and does not exhibit any response. This implies that the TLNs for the Schwarzschild black hole vanish, i.e., $\kappa_{\ell}^{s}$ = 0 \cite{Schwarzschild, Kol:2011vg, Hui:2020xxx}. It has been demonstrated that TLNs, in fact, vanish for several families of asymptotically flat black holes in four-dimensional GR \cite{LeTiec:2020spy, Kerr,Cardoso:2017cfl, Pereniguez:2021xcj, Rai:2024lho, Ma:2024few}. In Eq. (\ref{TLNs}), TLNs are defined as the ratio of induced response over the external source, as is the case in Newtonian Gravity. However, one should be careful regarding using the Newtonian matching procedure due to the inherent non-linearity of GR. In fact, since $\ell$ is an integer, the growing and decaying modes can mix, leading to an ambiguity in extracting TLNs. To overcome this ambiguity, one possible way is to analytically continue $\ell$, through which the growing and decaying modes decouple and TLNs can be defined uniquely, as it was discussed in Ref. \cite{LeTiec:2020spy}.

TLNs defined in Eq. (\ref{TLNs}) are obtained from the asymptotic expansion of linear static responses at large distances that are power-law in $\tilde{r}$. The coefficient is constant and independent of scale. However, in the presence of matter or backgrounds that deviate from GR, it is not guaranteed that the asymptotic expansion of the radial component of perturbations takes simple power-law series. Instead, the asymptotic behaviors can include logarithmic terms at large distances \cite{Katagiri:2023umb}
\bq \label{TLNs-log}
\Psi(r) \sim  \tilde{r}^{\ell+1} \left[1 + \mathcal{O}\left(\tilde{r}^{-1}\right)\right] + \kappa_{\ell}^{s}  \tilde{r}^{-\ell} \left[\ln \left(\tilde{r}\right) + \mathcal{O}(1)\right]\left[1 + \mathcal{O}\left(\tilde{r}^{-1}\right)\right].
\eq
The prefactor of the linear response term $\tilde{r}^{-\ell}$ includes a logarithmic term, and indeed it is scale dependent. The coefficient of the logarithmic term is interpreted as a classical renormalization group as discussed in Ref. \cite{Hui:2020xxx}. It has also been found that TLNs exhibit such logarithmic running for large multipole numbers, i.e., $\ell\geq 4$, in extended theories of gravity \cite{Barura:2024uog, Katagiri:2023umb}.

It should be noted that it is not always possible to transform Eq. (\ref{perturbation-r}) into a suitable form to find an exact solution. In this case, one can use the perturbative approach, assuming the smallness of the free parameter of the model. For the effective metrics considered in this manuscript, the quantum parameters are naturally small. Hence, it is useful to discuss how to extract TLNs in such cases when the exact solution is not obtainable due to the complicated form of metric functions. For this, we review the Green's function technique used in Ref. \cite{Barura:2024uog}, through which one can find TLNs up to arbitrary order unambiguously. It is useful to perform a change of variables such that all the modifications from GR are absorbed into the new effective potential. This can be obtained from the following field transformation
\begin{align}
    \Psi(\tilde{r}) = \frac{\Phi(\tilde{r})}{\sqrt{Z(\tilde{r})}},  \ \ \ \ \ \ Z(\tilde{r}) = \frac{\sqrt{A(\tilde{r})B(\tilde{r})}}{f(\tilde{r})}.
\end{align}
which transforms Eq. (\ref{perturbation-r}), in the static limit ($\omega=0$) and written in dimensionless quantity $\tilde{r}$, into the following form
\bq\label{perturbation-transformed}
f(\tilde{r}) \frac{\mathrm{d}}{\mathrm{d}\tilde{r}}\left(f(\tilde{r})\frac{\mathrm{d}}{\mathrm{d}\tilde{r}}\Phi(\tilde{r})\right)  - f(\tilde{r}) U(\tilde{r}) \Phi(\tilde{r}) = 0.
\eq
Here $f(\tilde{r})$ is the metric function for Schwarzschild black holes in four dimensions, i.e., $f(r) = 1-1/\tilde{r}$, and the new effective potential $U(\tilde{r})$ reads as
\bq
U(\tilde{r}) = \sqrt{\frac{A(\tilde{r})}{B(\tilde{r})}}  \frac{V(\tilde{r})}{Z(\tilde{r})} - \frac{1}{4 Z(\tilde{r})^{2}}\left[ f(\tilde{r}) \left(\frac{\mathrm{d} Z(\tilde{r})}{\mathrm{d}\tilde{r}}\right)^2 - 2 Z(\tilde{r}) \frac{\mathrm{d}}{\mathrm{d}\tilde{r}}\left(f(\tilde{r}) \frac{\mathrm{d} Z(\tilde{r})}{\mathrm{d}\tilde{r}}\right)\right].
\eq
For $A(\tilde{r}) = B(\tilde{r}) = f(\tilde{r})$, $Z(\tilde{r})=1$ and the effective potential $U(\tilde{r})$ reduces to the effective potential for the Schwarzschild black hole, i.e., $U(\tilde{r}) = V_{\textrm{GR}}(\tilde{r})$. To uniquely extract TLNs, we begin by treating the multipole number $\ell$ as a general complex number rather than an integer. Specifically, we introduce a small complex deviation, i.e., $\epsilon \in \mathbb{C}$, with $|\epsilon| \ll 1$, and consider $\ell \to \ell + \epsilon$. This approach allows for an analytic continuation of the solution near integer values of $\ell$. After solving the relevant equations in this extended setting, we then take the limit $\epsilon \to 0$ to recover the physical (integer) value of the multipole number. This ensures that the growing and decaying modes do not overlap, and the extracted TLNs are unique. 

From Eq. (\ref{perturbation-transformed}), one finds that it is similar to the equation for the Schwarzschild black hole with a new effective potential $U(\tilde{r})$ that encodes all the information of the modified metric. Let us consider the following expansion of the effective potential:
\bq\label{zero-order-potential}
U(\tilde{r}) = \sum_{k\ge 0} \eta^{k} U^{(k)}(\tilde{r}),  \ \ \ \ \ \ \ U^{(0)}(\tilde{r}) = V_{\textrm{GR}} (\tilde{r}),
\eq
where the parameter $\eta$ represents the deviation from GR. One can obtain the perturbative solution using the Green's function technique used in Ref. \cite{Barura:2024uog} by considering the following solution
\begin{align}
    \Phi(\tilde{r}) = \sum_{k\ge 0} \eta^{k} \Phi^{(k)}(\tilde{r}).
\end{align}
The strategy is then to solve Eq. (\ref{perturbation-transformed}) order by order in $\eta$. At zeroth order, $\mathcal{O} (\eta^{0})$, we have the standard equation for the Schwarzschild black hole in four dimensions
\bq\label{order-zero}
f(\tilde{r}) \frac{\mathrm{d}}{\mathrm{d}\tilde{r}}\left(f(\tilde{r})\frac{\mathrm{d}}{\mathrm{d}\tilde{r}}\Phi^{(0)}(\tilde{r})\right)  - f(\tilde{r}) U^{(0)}(\tilde{r}) \Phi^{(0)}(\tilde{r}) = 0.
\eq
Performing the following field transformation
\begin{align} 
    u = \frac{1}{\tilde{r}},  \ \ \ \ \ \ \ \ \Phi^{(0)}(\tilde{r}) = u^{\ell}(\tilde{r}) W(u(\tilde{r})),
\end{align}
one can transform Eq. (\ref{order-zero}) into the following hypergeometric equation
\begin{align}
    u(1-u)\frac{\mathrm{d}^2 W(u)}{\mathrm{d}u^2} + \left[c - (a+b+1)u\right]\frac{\mathrm{d}W(u)}{\mathrm{d}u}- a b W(u)=0,
\end{align}
with the parameters
\begin{align}
    a = \ell+1+s,  \ \ \ \ b = \ell+1-s,  \ \ \ \ \ c = 2 \ell+2.
\end{align}
The two independent solutions to this equation are written in terms of the hypergeometric function as
\begin{align}
    \Phi^{(0)}_{+}(\tilde{r}) &= \tilde{r}^{\ell+1} \, _2F_1\left(-\ell-s,-\ell + s;-2 \ell;\frac{1}{\tilde{r}}\right) , \label{zero-order-solutions-1}\\      \Phi^{(0)}_{-}(\tilde{r}) & = \tilde{r}^{-\ell}\, _2F_1\left(\ell+1-s,\ell+1+s;2 \ell+2;\frac{1}{\tilde{r}}\right),\label{zero-order-solutions-2}
\end{align}
where the coefficients are normalized to unity without losing the generality. For a generic complex $\ell$, the expansion of the solution at the horizon is divergent for both $\Phi^{(0)}_{+}(\tilde{r})$ and $\Phi^{(0)}_{-}(\tilde{r})$. However, one can find a horizon-regular solution in such a way that the logarithmic terms cancel each other at the horizon
\begin{align}\label{phi-zero}
\Phi^{(0)} (\tilde{r}) = \Phi^{\textrm{hor-reg}} (\tilde{r}) = \Phi^{(0)}_{+}(\tilde{r}) + \kappa_{\ell}^{s(0)} \Phi^{(0)}_{-}(\tilde{r}) .
\end{align}
Here $\kappa_{\ell}^{s(0)}$ is a constant given by
\begin{align}
    \kappa_{\ell}^{s(0)} = - \frac{\Gamma(-2\ell)\Gamma(\ell+1+s)\Gamma(\ell+1-s)}{\Gamma(-\ell-s)\Gamma(-\ell+s) \Gamma(2\ell+2)},
\end{align}
which goes to zero for physical multipoles $\ell \in \mathbb{Z}$. This is related to the fact that TLNs are vanishing for four-dimensional black holes in classical GR. Given the solution for the zeroth order equation, let's consider the $k$ order, $\mathcal{O}(\eta^{k})$, equation, which can be written as follows
\bq\label{order-k}
f(\tilde{r}) \frac{\mathrm{d}}{\mathrm{d}\tilde{r}}\left(f(\tilde{r})\frac{\mathrm{d}}{\mathrm{d}\tilde{r}}\Phi^{(k)}(\tilde{r})\right)  - f(\tilde{r}) U^{(0)}(\tilde{r}) \Phi^{(k)}(\tilde{r}) = f(\tilde{r}) \mathcal{S}^{(k)}(\tilde{r}),
\eq
with the source term $\mathcal{S}^{(k)}(\tilde{r})$ reading as 
\begin{align}\label{source-term}
\mathcal{S}^{(k)}(\tilde{r}) = \sum_{i= 0}^{k-1} U^{(k-i)}(\tilde{r}) \Phi^{(i)}(\tilde{r}).
\end{align}
Note that the LHS of Eq. (\ref{order-k}) is the same as the LHS of Eq. (\ref{order-zero}) with an extra source term in the RHS. So the homogeneous solution is a linear combination of $\Phi^{(0)}_{\pm}(\tilde{r})$, and we can find the inhomogeneous solution using the Green's function technique. We consider  the Green's function $G(\tilde{r}, \tilde{r}^{\prime})$, which satisfies the following equation
\bq\label{order-k-green}
\left[f(\tilde{r}) \frac{\mathrm{d}}{\mathrm{d}\tilde{r}}\left(f(\tilde{r})\frac{\mathrm{d}}{\mathrm{d}\tilde{r}}\right)  - f(\tilde{r}) U^{(0)}(\tilde{r})\right] G(\tilde{r}, \tilde{r}^{\prime}) = f(\tilde{r}) \delta(\tilde{r}-\tilde{r}^{\prime}),
\eq
with $\delta(\tilde{r})$ being the delta function. Imposing the regularity of solution at the horizon and also $\Phi(\tilde{r}) = \mathcal{O}(\tilde{r}^{\ell})$ as $\tilde{r} \rightarrow \infty$, the Green's function can be defined as (see appendix A in Ref. \cite{Barura:2024uog} for details of derivation)
\begin{align}
    G(\tilde{r}, \tilde{r}^{\prime}) = - \frac{1}{2\ell+1} \left[\Phi^{(0)}_{-}(\tilde{r}) \Phi^{\textrm{hor-reg}}(\tilde{r}^{\prime}) \Theta(\tilde{r}-\tilde{r}^{\prime}) +  \Phi^{\textrm{hor-reg}}(\tilde{r}) \Phi^{(0)}_{-}(\tilde{r}^{\prime}) \Theta(\tilde{r}^{\prime}-\tilde{r}) \right],
\end{align}
$\Theta(\tilde{r})$ is the Heaviside step function. One can then find the inhomogeneous solution at the $k$th order as follows
\begin{align}\label{solution-k}
  \nonumber  \Phi^{(k)}(\tilde{r}) &= \int_{1}^{\infty} G(\tilde{r}, \tilde{r}^{\prime}) \mathcal{S}^{(k)}(\tilde{r}^{\prime}) \mathrm{d}\tilde{r}^{\prime}\\ &=
    -\frac{1}{2\ell+1} \left[\Phi^{(0)}_{-}(\tilde{r}) \int_{1}^{\tilde{r}} \Phi^{\textrm{hor-reg}}(\tilde{r}^{\prime}) \mathcal{S}^{(k)}(\tilde{r}^{\prime}) \mathrm{d}\tilde{r}^{\prime} - \Phi^{\textrm{hor-reg}}(\tilde{r}) \int_{\infty}^{\tilde{r}} \Phi^{(0)}_{-}(\tilde{r}^{\prime}) \mathcal{S}^{(k)}(\tilde{r}^{\prime}) \mathrm{d}\tilde{r}^{\prime} \right].
\end{align}
Given this solution, one can expand the horizon-regular solution at large distances to extract TLNs at $k$ order. This analysis has been done in Ref. \cite{Barura:2024uog}, and it was found that for $\ell \in \mathbb{Z}$, TLNs are obtained from the constant part of the following integration
\begin{align}\label{TLNs-k}
    I[\mathcal{S}^{(k)}(\tilde{r})] = - \frac{1}{2\ell+1} \int_{1}^{\tilde{r}} \Phi^{(0)}_{+}(\tilde{r}^{\prime}) \mathcal{S}^{(k)}(\tilde{r}^{\prime}) \mathrm{d}\tilde{r}^{\prime}. 
\end{align}
However, we should point out that if the integration contains the logarithmic behavior, which is typically the case for a background geometry deviating from GR, that logarithmic term is also considered as part of TLNs. Given the solution at zeroth order and the effective potential containing modifications to GR, one can find the $k$ order solution and the corresponding TLNs. This is the procedure that we will use to compute the quantum gravitational corrections to the TLNs for covariant loop quantum black holes in section \ref{Section IV}.

\section{ Tidal Love numbers for Covariant Loop Quantum Black Holes}\label{Section IV}

In this section, we study the linear static response of covariant loop quantum black holes, discussed in Section \ref{Section II}, to an external tidal spin $s$ field, namely a scalar field, a vector field, and an axial gravitational field. Subsequently, we construct the horizon-regular solution perturbatively by applying the Green's function technique outlined in Section \ref{Section III} and extract the TLNs order by order. To this end, we first compute the TLNs for the ZLMY-I model in Section \ref{Section IV-A}, then for the ZLMY-II model in Section \ref{Section IV-B}, and finally present the results for the ABV model in Section \ref{Section IV-C}. 

\subsection{ZLMY-I model}\label{Section IV-A}

Here, we extract the TLNs for the ZLMY-I loop quantum black hole model \cite{Zhang:2024khj, Zhang:2024ney} in response to all three scalar, vector, and axial gravitational field perturbations. For this model, the metric functions, $A(r)$ and $B(r)$, in terms of dimensionless quantities can be written as follows
\bq\label{model-I-dimensionless}
A(\tilde{r}) = B(\tilde{r}) = f(\tilde{r}) \left[1+  \frac{{\tilde{\xi}}^2}{\tilde{r}^2}f(\tilde{r})\right], \  \ \ \ \ \ \ \ \ \ f(\tilde{r}) = 1 - \frac{1}{\tilde{r}}.
\eq
We have here introduced dimensionless quantum parameter, $\tilde{\xi} = \xi / r_H = \xi / (2M)$, which is a small number for macroscopic black holes. To find a perturbative solution for Eq.~(\ref{perturbation-r}) in the static limit, we consider $\tilde{\xi}^2$ as the small parameter that governs the perturbation series expansion. Specifically, the perturbative solution is constructed as a series expansion in powers of $\tilde{\xi}^2$, where the term of the first order in the expansion corresponds to $\mathcal{O}(\tilde{\xi}^2)$. For metric functions (\ref{model-I-dimensionless}), one can transform Eq. (\ref{perturbation-r}), in the static limit, into Eq. (\ref{perturbation-transformed}) by the following transformation
\begin{align}
    \Psi(\tilde{r}) =  \frac{\Phi(\tilde{r})}{\sqrt{Z(\tilde{r})}},  \ \ \ \ \ \ \ Z(\tilde{r}) = \left[1+  \frac{\tilde{\tilde{\xi}}^2}{\tilde{r}^2}f(\tilde{r})\right].
\end{align}
Subsequently, the zeroth order effective potential $U^{(0)}(\tilde{r})$ is given by Eq. (\ref{zero-order-potential}), and the first, second, and third order effective potentials read as
\begin{align}
U^{(1)}_{\textrm{ZLMY-I}}(\tilde{r})& =\frac{-\ell(\ell+1)+s(s+1)+1}{\tilde{r}^4} + \frac{\ell(\ell+1)-s (s+4)-5}{\tilde{r}^5} +  \frac{6 s+9}{2 \tilde{r}^6} \label{U1-model-I},\\
\nonumber U_{\textrm{ZLMY-I}}^{(2)}(\tilde{r}) & =  \frac{\ell(\ell+1)-s(s+1)-2}{\tilde{r}^6} +\frac{-2 \ell(\ell+1) + s(2s+5) +10}{\tilde{r}^7} \\ &+ \frac{4 \ell(\ell+1)-4 s(s+7) -59}{4 \tilde{r}^8} + \frac{3(4s+9)}{4 \tilde{r}^9},\\
\nonumber U_{\textrm{ZLMY-I}}^{(3)}(\tilde{r}) & =  \frac{-\ell(\ell+1)+s(s+1)+3}{\tilde{r}^8} + \frac{3 \ell(\ell+1)-3s (s+2) -17}{\tilde{r}^9}  \\ &  + \frac{-3 \ell(\ell+1)+3 s(s+4) +34}{ \tilde{r}^{10}}  + \frac{ \ell(\ell+1)- s(s+10) -29}{\tilde{r}^{11}} + \frac{3(s+3)}{\tilde{r}^{12}}.
\end{align}
From above equations, one finds that the first order modification to the effective potential has corrections from $\mathcal{O}(\tilde{r}^{-4})$ up to $\mathcal{O}(\tilde{r}^{-6})$, while it has corrections from $\mathcal{O}(\tilde{r}^{-6})$ to $\mathcal{O}(\tilde{r}^{-9})$ and from $\mathcal{O}(\tilde{r}^{-8})$ to $\mathcal{O}(\tilde{r}^{-12})$ for the second and third order modifications, respectively. Given the perturbative modifications to the effective potential and the zeroth order solutions (\ref{zero-order-solutions-1}) and (\ref{zero-order-solutions-2}), one can find the higher order solutions and then extract the TLNs for all three scalar, vector, and axial gravitational field perturbations order by order.

\subsubsection{Scalar field response}

For the scalar field perturbation, $s=0$, and the lowest multipole number, $\ell=0$, the growing and decaying modes at the zeroth order, $\mathcal{O}(\tilde{\xi}^{0})$, which are given by Eqs. (\ref{zero-order-solutions-1}) and (\ref{zero-order-solutions-2}), take the following form
\begin{align}
    \Phi^{(0)}_{+}(\tilde{r}) &= \Phi^{(0)}(\tilde{r}) = \Phi^{\textrm{hor-reg}}(\tilde{r}) = \tilde{r}, \label{solution-s=0-l=0-1} \\  \Phi^{(0)}_{-}(\tilde{r}) &= -\tilde{r} \log \left(1-\frac{1}{\tilde{r}}\right).\label{solution-s=0-l=0-2}
\end{align}
We should point out that the first equality in the first equation comes from Eq. (\ref{phi-zero}) and is due to the fact that $\kappa_{\ell}^{s}$ is zero for the integer multipole number $\ell$ in the case of the classical Schwarzschild black hole. For $s=0$ and $\ell=0$, at the first order, $\mathcal{O} (\tilde{\xi}^2)$, the effective potential and the source term can be found from Eqs. (\ref{U1-model-I}) and (\ref{source-term}), respectively
\begin{align}
    U^{(1)}(\tilde{r}) &= \frac{9}{2 \tilde{r}^6}-\frac{5}{\tilde{r}^5}+\frac{1}{\tilde{r}^4},  \\ \mathcal{S}^{(1)}(\tilde{r}) &= U^{(1)}(\tilde{r}) \Phi^{(0)}(\tilde{r}) = \frac{9}{2 \tilde{r}^5}-\frac{5}{\tilde{r}^4}+\frac{1}{\tilde{r}^3}.
\end{align}
Having the first order source term and $\Phi^{(0)}_{+}(\tilde{r})$ given by Eq. (\ref{solution-s=0-l=0-1}), the TLN at the first order can be extracted from the constant part of the following integration
\begin{align}
    I[\mathcal{S}^{(1)}(\tilde{r})] = - \int_{1}^{\tilde{r}} \Phi^{(0)}_{+}(\tilde{r}^{\prime}) \mathcal{S}^{(1)}(\tilde{r}^{\prime}) \mathrm{d} \tilde{r}^{\prime} = \frac{3}{2 \tilde{r}^3}-\frac{5}{2 \tilde{r}^2}+\frac{1}{\tilde{r}}.
\end{align}
This implies that the first order correction to the TLN for $s=0$ and $\ell=0$ is zero, i.e., $\kappa_{0}^{0(1)} =0$. Moreover, the first order solution can be computed as follows
\begin{align}
    \Phi^{(1)} (\tilde{r}) = \int_{1}^{\infty} G(\tilde{r}, \tilde{r}^{\prime}) \mathcal{S}^{(1)}(\tilde{r}^{\prime}) \textrm{d}\tilde{r}^{\prime} =  -\frac{1}{2 \tilde{r}^2} + \frac{1}{2 \tilde{r}}.
\end{align}
From the first order solution, one can build the second order source term. Hence, the second order effective potential and source term read as
\begin{align}
    U^{(2)}(\tilde{r}) &=\frac{27}{4 \tilde{r}^9}-\frac{59}{4 \tilde{r}^8}+\frac{10}{\tilde{r}^7}-\frac{2}{\tilde{r}^6}, \\  \mathcal{S}^{(2)}(\tilde{r}) &= U^{(2)}(\tilde{r}) \Phi^{(0)}(\tilde{r}) + U^{(1)}(\tilde{r}) \Phi^{(1)}(\tilde{r})  =  \frac{9}{2 \tilde{r}^8}-\frac{10}{\tilde{r}^7}+\frac{7}{\tilde{r}^6}-\frac{3}{2 \tilde{r}^5},
\end{align}
from which one can read the second order correction to the TLN from the following integration
\begin{align}
    I[\mathcal{S}^{(2)}(\tilde{r})] = - \int_{1}^{\tilde{r}} \Phi^{(0)}_{+}(\tilde{r}^{\prime}) \mathcal{S}^{(2)}(\tilde{r}^{\prime}) \mathrm{d} \tilde{r}^{\prime} = \frac{3}{4 \tilde{r}^6}-\frac{2}{\tilde{r}^5}+\frac{7}{4 \tilde{r}^4}-\frac{1}{2 \tilde{r}^3},
\end{align}
indicating that $\kappa_{1}^{0(2)}=0$. Then, one can find the following second order solution
\begin{align}
    \Phi^{(2)}(\tilde{r}) = \int_{1}^{\infty} G(\tilde{r}, \tilde{r}^{\prime}) \mathcal{S}^{(2)}(\tilde{r}^{\prime}) \mathrm{d}\tilde{r}^{\prime} = -\frac{1}{8 \tilde{r}^5}+\frac{1}{4 \tilde{r}^4}-\frac{1}{8 \tilde{r}^3}.
\end{align}
We can next write the third order effective potential and source term as follows
\begin{align}
    U^{(3)}(\tilde{r}) &=\frac{9}{\tilde{r}^{12}}-\frac{29}{\tilde{r}^{11}}+\frac{34}{\tilde{r}^{10}}-\frac{17}{\tilde{r}^9}+\frac{3}{\tilde{r}^8}, \\ \nonumber \mathcal{S}^{(3)}(\tilde{r}) &= U^{(3)}(\tilde{r}) \Phi^{(0)}(\tilde{r}) + U^{(2)}(\tilde{r})\Phi^{(1)}(\tilde{r}) + U^{(1)}(\tilde{r}) \Phi^{(2)}(\tilde{r})  \\ & =  \frac{81}{16 \tilde{r}^{11}}-\frac{33}{2 \tilde{r}^{10}}+\frac{315}{16 \tilde{r}^9}-\frac{81}{8 \tilde{r}^8}+\frac{15}{8 \tilde{r}^7}.
\end{align}
Hence, the third order correction to the TLN can be obtained from the constant part of the following integration
\begin{align}
 I[\mathcal{S}^{(3)}(\tilde{r})] &= - \int_{1}^{\tilde{r}} \Phi^{(0)}_{+}(\tilde{r}^{\prime}) \mathcal{S}^{(3)}(\tilde{r}^{\prime}) \mathrm{d} \tilde{r}^{\prime}  = \frac{9}{16 \tilde{r}^9}-\frac{33}{16 \tilde{r}^8}+\frac{45}{16 \tilde{r}^7}-\frac{27}{16 \tilde{r}^6}+\frac{3}{8 \tilde{r}^5}.
\end{align}
This implies that the TLN in response to a scalar field and for $\ell=0$ up to the third order, $\mathcal{O}(\tilde{\xi}^6)$, vanishes, i.e., $\kappa_{0}^{0}=0$. This result is in agreement with AOS model, which predicts a vanishing TLN only for scalar field response with $\ell=0$ \cite{Motaharfar:2025typ}.

For $\ell=1$, the zeroth order solutions (\ref{zero-order-solutions-1}) and (\ref{zero-order-solutions-2}) take the following form
\begin{align}
    \Phi^{(0)}_{+}(\tilde{r}) & = \Phi^{(0)}(\tilde{r}) = \Phi^{\textrm{hor-reg}}(\tilde{r}) =  \tilde{r}^2-\frac{\tilde{r}}{2}, \label{solution-s=0-l=1-1} \\ \Phi^{(0)}_{-}(\tilde{r}) & = -12\left[  \left( \tilde{r}^2 - \frac{\tilde{r}}{2}\right) \log \left(1-\frac{1}{\tilde{r}}\right) + \tilde{r} \right].\label{solution-s=0-l=1-2}
\end{align}
Inserting the first order effective potential (\ref{U1-model-I}) and the zeroth order solution (\ref{solution-s=0-l=1-1}) into Eq. (\ref{source-term}), one can find the first order source term from which the following integration can be expressed as 
\begin{align}
    I[\mathcal{S}^{(1)}(\tilde{r})] = -\frac{1}{3} \int_{1}^{\tilde{r}} \Phi^{(0)}_{+}(\tilde{r}^{\prime}) \mathcal{S}^{(1)}(\tilde{r}^{\prime}) \mathrm{d} \tilde{r}^{\prime} = \frac{1}{8 \tilde{r}^3}-\frac{7}{8 \tilde{r}^2}+\frac{29}{12 \tilde{r}}-2+\frac{2 \log (\tilde{r})}{3} +\frac{\tilde{r}}{3},
\end{align}
denoting that $\kappa_{1}^{0(1)} = -2 + 2\log(\tilde{r})/3$. So the scalar TLN for $\ell=1$ is negative and has a logarithmic running at the leading order in contrast to $\ell=0$ with a vanishing TLN. This is also in contrast to the AOS model, which does not exhibit logarithmic running for all multipole numbers and all three scalar, vector, and axial gravitational field responses \cite{Motaharfar:2025typ}. Computing the TLN up to the second order, we find
\begin{align}
\kappa_{1}^{0} &= \left[-2\tilde{\xi}^2  -\left(\frac{557+ 192 \pi ^2}{36}\right) \tilde{\xi}^4 + \mathcal{O}(\tilde{\xi}^6) \right] + \left[\frac{2}{3}\tilde{\xi}^2 + \left(\frac{16 \pi ^2-24}{9}\right)\tilde{\xi}^4 +\mathcal{O}(\tilde{\xi}^6)  \right]\log \left(\tilde{r}\right),
\end{align}
from which one can observe that the second order solution negatively effects the TLN, while positively contributing to the running. We should point out that we just compute the TLN up to the second order since the integration (\ref{solution-k}) cannot be done analytically, whereby we cannot construct the third order source term. 

Similarly, for $\ell=2$, the growing and decaying modes at the zeroth order are given by
\begin{align}
    \Phi^{(0)}_{+}(\tilde{r}) &= \Phi^{(0)}(\tilde{r}) =  \Phi^{\textrm{hor-reg}}(\tilde{r}) = \tilde{r}^3-\tilde{r}^2+\frac{\tilde{r}}{6}, \label{solution-s=0-l=2-1}  \\ \Phi^{(0)}_{-}(\tilde{r}) &= -180\left[  \left( \tilde{r}^3-\tilde{r}^2+\frac{\tilde{r}}{6}\right) \log \left(1-\frac{1}{\tilde{r}}\right) + {\tilde{r}^2} - \frac{\tilde{r}}{2} \right],\label{solution-s=0-l=2-2}
\end{align}
and the first order correction to the TLN can be extracted from the following integration
\begin{align}
    \nonumber I[\mathcal{S}^{(1)}(\tilde{r})] &= -\frac{1}{5} \int_{1}^{\tilde{r}} \Phi^{(0)}_{+}(\tilde{r}^{\prime}) \mathcal{S}^{(1)}(\tilde{r}^{\prime}) \mathrm{d} \tilde{r}^{\prime} \\ & = \frac{1}{120 \tilde{r}^3}-\frac{53}{360 \tilde{r}^2}+\frac{199}{180 \tilde{r}}-\frac{31}{30}+\frac{6 \log (\tilde{r})}{5} +\frac{5 \tilde{r}}{6}-\frac{11 \tilde{r}^2}{10}+\frac{\tilde{r}^3}{3}.
\end{align}
This implies that the first order correction to the TLN is $\kappa_{2}^{0(1)} = -31/30 + 6\log(\tilde{r})/5$, which is negative and has a positive running similar to $\ell=1$. One can then compute the TLN up to the second order as follows
\begin{align}
\kappa_{2}^{0} &= \left[-\frac{31}{30} \tilde{\xi}^2 - \left(\frac{11276 +558 \pi^2}{15}\right)\tilde{\xi}^4 + \mathcal{O}(\tilde{\xi}^6) \right] + \left[\frac{6 }{5} \tilde{\xi}^2 + \left(\frac{216 \pi ^2- 426}{5}\right)\tilde{\xi}^4 + \mathcal{O}(\tilde{\xi}^6) \right]\log \left(\tilde{r}\right),
\end{align}
demonstrating that the second order solution negatively impacts the TLN and positively contributes to the running, similar to $\ell=1$. In summary, we found that the TLNs are negative in response to the scalar field perturbation in the case of the ZLMY-I loop quantum black hole model at the leading order, $\mathcal{O}(\tilde{\xi}^2)$, for $\ell=1$ and $\ell=2$, while it vanishes for $\ell=0$. Moreover, our results show that the TLNs exhibit positive logarithmic running at the leading order even for low multipole numbers, i.e., $\ell=1$ and $\ell=2$.

\subsubsection{Vector field response}

Now we compute the vector TLNs for the ZLMY-I model. For $s=1$ and $\ell=1$, the zeroth order growing and decaying solutions are given by
\begin{align}
    \Phi^{(0)}_{+}(\tilde{r}) & = \Phi^{(0)}(\tilde{r})= \Phi^{\textrm{hor-reg}}(\tilde{r}) =  \tilde{r}^2, \label{solution-s=1-l=1-1} \\  \Phi^{(0)}_{-}(\tilde{r}) &= -3 \left[\tilde{r}^2 \log \left(1-\frac{1}{\tilde{r}}\right)+ \tilde{r}+\frac{1}{2}\right].\label{solution-s=1-l=1-2}
\end{align}
The first order correction to the TLN can also be obtained from the following integration
\begin{align}
    I[\mathcal{S}^{(1)}(\tilde{r})] = -\frac{1}{3} \int_{1}^{\tilde{r}} \Phi^{(0)}_{+}(\tilde{r}^{\prime}) \mathcal{S}^{(1)}(\tilde{r}^{\prime}) \mathrm{d} \tilde{r}^{\prime} = \frac{5}{2 \tilde{r}}-\frac{13}{6} +\frac{8 \log (\tilde{r})}{3} -\frac{\tilde{r}}{3},
\end{align}
which turns out to be $\kappa_{1}^{1(1)} = -{13}/{6} + {8 \log (\tilde{r})}/{3}$. This TLN is negative and has a positive logarithmic running. Although the negative sign of the TLN is similar to that predicted by the AOS model, the presence of running introduces a distinct feature. Straightforwardly, the TLN up to the second order can be extracted as follows
\begin{align}
\kappa_{1}^{1} &= \left[-\frac{13}{6} \tilde{\xi}^2 -\left(\frac{437+52 \pi ^2}{9}\right) \tilde{\xi}^4 + \mathcal{O}(\tilde{\xi}^6)  \right] + \left[\frac{8 }{3}\tilde{\xi}^2 + \left(\frac{32\pi ^2 + 24}{9} \right)\tilde{\xi}^4 + \mathcal{O}(\tilde{\xi}^6)\right]\log \left(\tilde{r}\right).
\end{align}
From which one observes that the second order contribution to the TLN is negative, while the contribution to the running is positive.

For $\ell=2$, the growing and decaying modes at the zeroth order read as
\begin{align}
    \Phi^{(0)}_{+}(\tilde{r}) &=  \Phi^{(0)}(\tilde{r}) =  \Phi^{\textrm{hor-reg}}(\tilde{r}) = \tilde{r}^3-\frac{3}{4} \tilde{r}^{2}, \label{solution-s=1-l=2-1}  \\ \Phi^{(0)}_{-}(\tilde{r}) &= -80\left[  \left( \tilde{r}^3-\frac{3}{4} \tilde{r}^{2}\right) \log \left(1-\frac{1}{\tilde{r}}\right) + \tilde{r}^2 - \frac{r}{4} - \frac{1}{24}\right],\label{solution-s=1-l=2-2}
\end{align}
and the integration for extracting the first order correction to the TLN is given by 
\begin{align}
    I[\mathcal{S}^{(1)}(\tilde{r})] = -\frac{1}{5} \int_{1}^{\tilde{r}} \Phi^{(0)}_{+}(\tilde{r}^{\prime}) \mathcal{S}^{(1)}(\tilde{r}^{\prime}) \mathrm{d} \tilde{r}^{\prime} = \frac{27}{32 \tilde{r}} +\frac{219}{160} +\frac{27 \log (\tilde{r})}{10} -\frac{189 \tilde{r}}{80} -\frac{\tilde{r}^2}{20} + \frac{\tilde{r}^3}{5}.
\end{align}
Hence, the first order correction to the TLN is $\kappa_{2}^{1(1)} = {219}/{160} +{27}\log (\tilde{r})/{10}$, which is positive and has a positive running at the leading order. This is in contrast to the AOS model, which predicts a negative TLN without running for $\ell=2$. Subsequently, the TLN for $\ell=2$ up to the second order turns out to be
\begin{align}
\kappa_{2}^{1} &= \left[\frac{219}{160} \tilde{\xi}^2 + \left(\frac{1971 \pi ^2}{40}-\frac{14289}{16}\right)\tilde{\xi}^4 + \mathcal{O}(\tilde{\xi}^6) \right] + \left[\frac{27}{10} \tilde{\xi}^2 + \left(\frac{5013 + 1944 \pi^2}{10}\right) \tilde{\xi}^4 + \mathcal{O}(\tilde{\xi}^6) \right]\log \left(\tilde{r}\right),
\end{align}
from which one can find that the second order solution negatively affects TLN, while positively contributes to its running. 

Finally, for $\ell=3$, we have
\begin{align}
    \Phi^{(0)}_{+}(\tilde{r}) &= \Phi^{(0)}(\tilde{r}) = \Phi^{\textrm{hor-reg}}(\tilde{r}) =  \tilde{r}^4-\frac{4 \tilde{r}^3}{3}+\frac{2 \tilde{r}^2}{5}, \label{solution-s=1-l=3-1}  \\ \Phi^{(0)}_{-}(\tilde{r}) &= -1575\left[\left(\tilde{r}^4-\frac{4 \tilde{r}^3}{3}+\frac{2 \tilde{r}^2}{5}\right) \log \left(1-\frac{1}{\tilde{r}}\right) + \tilde{r}^3-\frac{5 \tilde{r}^2}{6}+\frac{\tilde{r}}{15}+\frac{1}{180}\right].\label{solution-s=1-l=3-2}
\end{align}
We can similarly read the first order correction to the TLN from the following expression
\begin{align}
    \nonumber I[\mathcal{S}^{(1)}(\tilde{r})] &= -\frac{1}{7} \int_{1}^{\tilde{r}} \Phi^{(0)}_{+}(\tilde{r}^{\prime}) \mathcal{S}^{(1)}(\tilde{r}^{\prime}) \mathrm{d} \tilde{r}^{\prime} \\& = \frac{6}{35 \tilde{r}}+\frac{722}{525} +\frac{192 \log (\tilde{r})}{175} -\frac{394 \tilde{r}}{175} +\frac{118 \tilde{r}^2}{315} +\frac{631 \tilde{r}^3}{630}-\frac{13 \tilde{r}^4}{14} + \frac{9 \tilde{r}^5}{35}.
\end{align}
which is $\kappa_{3}^{1(1)} = 722/525 + 192 \log(\tilde{r})/175$. It is positive and has a positive running similar to $\ell=2$. This is again in contrast to the AOS model, which has a negative TLN without running behavior for $\ell=3$. Finally, the TLN up to the second order reads as
\begin{align}
\nonumber \kappa_{3}^{1} &= \left[\frac{722}{525} \tilde{\xi}^2 + \left(\frac{69312 \pi ^2}{175}-\frac{655553}{700}\right)\tilde{\xi}^4 + \mathcal{O}(\tilde{\xi}^6) \right] \\ &+ \left[\frac{192 }{175} \tilde{\xi}^2 + \left(\frac{578104 +55296 \pi ^2}{175}\right)\tilde{\xi}^4 + \mathcal{O}(\tilde{\xi}^6) \right]\log \left(\tilde{r}\right),
\end{align}
implying that the second order solution also negatively impacts the TLN and positively contributes to the running similar to $\ell=2$. To summarize, we found that the TLNs in response to vector field perturbation are negative at the leading order for $\ell=1$, while they have positive values for $\ell=2$ and $\ell=3$.  This is in contrast to scalar field responses, which are negative for $\ell=1$ and $\ell=2$ and vanishing for $\ell=0$. The vector TLNs also exhibits  positive logarithmic running at the leading order for the three lowest multipole numbers.

\begin{table}[t!]
\centering
\begin{tabular}{ c c c c c c c c c c}
\hline
\hline
 &   & $s=0$ &  &  & $s=1$ &  &  & $s=2$ &  \\ 
\hline
\ \ \ \ \ \  &  \ \ \  $\ell=0$  \ \ \  &  \ \ \  $\ell=1$  \ \ \  &  \ \ \  $\ell=2$  \ \ \  &  \ \ \  $\ell=1$  \ \ \ &  \ \ \  $\ell=2$  \ \ \  &  \ \ \  $\ell=3$  \ \ \  &  \ \ \  $\ell=2$  \ \ \  &  \ \ \ $\ell=3$  \ \ \  &  \ \ \  $\ell=4$  \ \ \  \\ 
\hline
$C(s, \ell)$ & $0$ & $-2.5847$ & $-1.3354 $ & $-1.0500$ & $1.7689$ & $1.7773 $ & $1.3785 $ & $0.3336$ & $0.0613$  \\ \hline
$C_{\textrm{log}}(s, \ell)$ &$0$ & $0.8616$ & $1.5508$ & $3.4462$ & $3.4893$ & $1.4179$ & $0$ & $0$ & $0$ \\
\hline
\hline
\end{tabular}
\caption{The leading order corrections to the TLNs and their running, i.e., Eq. (\ref{TLN-model-I-II}), for the ZLMY-I loop quantum black hole model and the three lowest multipole numbers $\ell$ in response to the scalar, vector, and axial gravitational field perturbations.}
\label{TLN-first-order-model-I}
\end{table}

\subsubsection{Axial gravitational field response}

Here, we extract the TLNs in response to the axial gravitational field perturbation for the ZLMY-I model. For $s=2$ and $\ell=2$, we have
\begin{align}
    \Phi^{(0)}_{+}(\tilde{r}) &= \Phi^{(0)}(\tilde{r}) = \Phi^{\textrm{hor-reg}}(\tilde{r}) =  \tilde{r}^3, \label{solution-s=2-l=2-1}  \\ \Phi^{(0)}_{-}(\tilde{r}) &= -5\left[\tilde{r}^3 \log \left(1-\frac{1}{\tilde{r}}\right)+\tilde{r}^2+\frac{\tilde{r}}{2}+\frac{1}{4 \tilde{r}}+\frac{1}{3}\right].\label{solution-s=2-l=2-2}
\end{align}
Hence, the integration for extracting the first order correction to the TLN reads as
\begin{align}
I[\mathcal{S}^{(1)}(\tilde{r})] = -\frac{1}{5} \int_{1}^{\tilde{r}} \Phi^{(0)}_{+}(\tilde{r}^{\prime}) \mathcal{S}^{(1)}(\tilde{r}^{\prime}) \mathrm{d} \tilde{r}^{\prime} = -\frac{\tilde{r}^3}{15}+\frac{11 \tilde{r}^2}{10}-\frac{21 \tilde{r}}{10}+\frac{16}{15},
\end{align}
denoting $\kappa_{2}^{2(1)} = 16/15$, which is positive and has no running. This is in contrast to the AOS model, which predicts a negative TLN in response to an axial gravitational field for $\ell=2$ with no running. Computing the TLN up to the third order, one finds that
\begin{align}
\kappa_{2}^{2} &= \left[\frac{16}{15} \tilde{\xi}^2 +  \frac{1027}{180} \tilde{\xi}^4 - \left(\frac{14534+1408 \pi ^2}{405}\right) \tilde{\xi}^6 + \mathcal{O}(\tilde{\xi}^8) \right] + \left[ -\frac{176}{45} \tilde{\xi}^4 -\frac{12452}{405} \tilde{\xi}^6 + \mathcal{O}(\tilde{\xi}^8) \right]\log (\tilde{r}).
\end{align}
From which one can find that the TLN exhibits a logarithmic running at the second order. Moreover, the second order solution positively effects the TLN, while negatively contributing to the running.

For $\ell=3$, the zeroth order growing and decaying solutions are given by
\begin{align}
    \Phi^{(0)}_{+}(\tilde{r}) &= \Phi^{(0)}(\tilde{r})= \Phi^{\textrm{hor-reg}}(\tilde{r}) =  \tilde{r}^4-\frac{5 \tilde{r}^3}{6}, \label{solution-s=2-l=3-1} \\ \Phi^{(0)}_{-}(\tilde{r}) &= -252\left[\left(\tilde{r}^4-\frac{5 \tilde{r}^3}{6}\right) \log \left(1-\frac{1}{\tilde{r}}\right) + \tilde{r}^3-\frac{\tilde{r}^2}{3}-\frac{\tilde{r}}{12}-\frac{1}{36}-\frac{1}{120 \tilde{r}}\right].\label{solution-s=2-l=3-2}
\end{align}
Subsequently, one can find the first order correction to the TLN from the following expression
\begin{align}
 I[\mathcal{S}^{(1)}(\tilde{r})] & = -\frac{1}{7} \int_{1}^{\tilde{r}} \Phi^{(0)}_{+}(\tilde{r}^{\prime}) \mathcal{S}^{(1)}(\tilde{r}^{\prime}) \mathrm{d} \tilde{r}^{\prime}  =\frac{\tilde{r}^5}{7}-\frac{5 \tilde{r}^4}{42}-\frac{79 \tilde{r}^3}{108}+\frac{755 \tilde{r}^2}{504}-\frac{25 \tilde{r}}{24}+\frac{95}{378},
\end{align}
implying that $\kappa_{3}^{2(1)} = 95/378$, which is again positive with no running similar to $\ell=2$. This is again in contrast to the AOS model with a negative TLN and no running for $\ell=3$. We can next compute the TLN up to the third order as follows
\begin{align}
\nonumber \kappa_{3}^{2} & = \left[\frac{95}{368} \tilde{\xi}^2 + \frac{38551}{9072} \tilde{\xi}^4 - \left(\frac{103483511}{81648}+\frac{361000 \pi ^2}{5103}\right)\tilde{\xi}^6 + \mathcal{O}(\tilde{\xi}^8) \right]  \\& + \left[-\frac{3800}{567} \tilde{\xi}^4 -\frac{4633910}{5103}\tilde{\xi}^6 + \mathcal{O}(\tilde{\xi}^8) \right]\log \left(\tilde{r}\right).
\end{align}
This shows that the TLN exhibits a logarithmic running at the second order. The second order solution also positively impacts the TLN and has a negative contribution for its running, similar to $\ell=2$.

Finally, for $\ell=4$, we have
\begin{align}
    \Phi^{(0)}_{+}(\tilde{r}) &= \Phi^{(0)}(\tilde{r}) = \Phi^{\textrm{hor-reg}}(\tilde{r}) =  \tilde{r}^5-\frac{3 \tilde{r}^4}{2}+\frac{15 \tilde{r}^3}{28}, \label{solution-s=2-l=4-1} \\ \Phi^{(0)}_{-}(\tilde{r}) &= -7056\left[\left(\tilde{r}^5-\frac{3 \tilde{r}^4}{2}+\frac{15 \tilde{r}^3}{28}\right) \log \left(1-\frac{1}{\tilde{r}}\right) + \tilde{r}^4-\tilde{r}^3+\frac{5 \tilde{r}^2}{42}+\frac{\tilde{r}}{56}+\frac{1}{280}+\frac{1}{1680 \tilde{r}}\right].\label{solution-s=2-l=4-2}
\end{align}
Then the first order correction to the TLN can be extracted from the following integration
\begin{align}
\nonumber I[\mathcal{S}^{(1)}(\tilde{r})] & = -\frac{1}{9} \int_{1}^{\tilde{r}} \Phi^{(0)}_{+}(\tilde{r}^{\prime}) \mathcal{S}^{(1)}(\tilde{r}^{\prime}) \mathrm{d} \tilde{r}^{\prime} \\ & = \frac{13 \tilde{r}^7}{63}-\frac{7 \tilde{r}^6}{9}+\frac{389 \tilde{r}^5}{420}+\frac{\tilde{r}^4}{56}-\frac{2293 \tilde{r}^3}{2352}+\frac{1395 \tilde{r}^2}{1568}-\frac{75 \tilde{r}}{224}+\frac{93}{1960},
\end{align}
which indicates $\kappa_{4}^{2(1)} =93/1960 $ and is again positive without running, similar to $\ell=2$ and $\ell=3$. This is again in contrast to the AOS model, predicting a negative TLN with no running for $\ell=4$ in response to an axial gravitational field perturbation. Finally, the TLN up to the third order can be found as follows
\begin{align}
\nonumber \kappa_{4}^{2} & =\left[ \frac{93}{1960} \tilde{\xi}^2 + \frac{30497}{47040} \tilde{\xi}^4 -  \left(\frac{1295657681}{168000}+\frac{3736368 \pi ^2}{6125}\right) \tilde{\xi}^6 + \mathcal{O}(\tilde{\xi}^8) \right] \\ & + \left[-\frac{6696} {1225} \tilde{\xi}^4 -\frac{148193404}{30625}\tilde{\xi}^6 + \mathcal{O}(\tilde{\xi}^8) \right]\log \left(\tilde{r}\right),
\end{align}
from which one can see that it exhibits a logarithmic running at the second order. Moreover, the second order solution positively contributes to the TLN and negatively effects the running, similar to $\ell=2$ and $\ell=3$. Therefore, our results for the axial gravitational field perturbation show that the leading order corrections to the TLNs are positive with no running for the three lowest multipole numbers. However, they exhibit negative running at the second order. 

Before closing this section, let's summarize the result found for the ZLMY-I model. Using the value of the area gap in LQG, one can find that at the leading order, the TLNs for the ZLMY-I loop quantum black hole model can be written in a compact form as follows (in Planck units)
\begin{align}\label{TLN-model-I-II}
    \kappa_{\ell}^{s} = \left[ C(\ell, s)  +  C_{\textrm{log}}(\ell, s)  \log(\tilde{r})\right] M^{-\alpha} ,
\end{align}
where $C(\ell, s)$ and $C_{\textrm{log}}(\ell, s)$ are constants fixed by the spin of the external tidal field and multipole number, and they can be either negative, positive, or zero, and $\alpha$ depends on the specified model. Interestingly, for the ZLMY-I model, we have $\alpha=2$, indicating that the TLNs are Planck scale suppressed. This means that the TLNs are small for astrophysical black holes, while tidal deformability becomes stronger once the black hole mass reaches Planck mass. Such Planck scale suppression is also found for the AOS model, while the suppression is weaker in that case, given by $\kappa_{\ell}^{s} \propto M^{-2/3}$. We record the value of $C(\ell, s)$ and $C_{\textrm{log}}(\ell, s)$ in Table \ref{TLN-first-order-model-I} for comparison. As is obvious from Table \ref{TLN-first-order-model-I}, considering the same mass, the largest magnitude for the TLNs is obtained for scalar field response with $\ell=1$. While the smallest one is for the axial gravitational field response with $\ell=4$. The largest magnitude for the running occurs for vector field response with $\ell=2$.

\subsection{ZLMY-II model}\label{Section IV-B}

In this section, we extract the TLNs for all three scalar, vector, and axial gravitational field responses considering the ZLMY-II loop quantum black hole model \cite{Zhang:2024khj, Zhang:2024ney}. Writing the metric functions in terms of dimensionless quantities, they take the following form
\begin{align}
A(\tilde{r}) = f(\tilde{r}) = 1- \frac{1}{\tilde{r}}, \ \ \ \ \ \ \ \ \ \  
B(\tilde{r})  =  f(\tilde{r}) \left[1+  \frac{\tilde{\xi}^2}{\tilde{r}^2}f(\tilde{r})\right].
\end{align}
We can transform Eq. (\ref{perturbation-r}), in the static limit, to Eq. (\ref{perturbation-transformed}) with the new effective potential containing the quantum gravitational effects using the following field redefinition
\begin{align}
   \Psi(\tilde{r}) =  \frac{\Phi(\tilde{r})}{\sqrt{Z(\tilde{r})}},  \ \ \ \ \ \ \   Z(r) = \sqrt{ 1 +  \frac{\tilde{\xi}^2}{\tilde{r}^2}f(\tilde{r})}.
\end{align}
While the effective potential up to the third order reads as
\begin{align}
U^{(1)}_{\textrm{ZLMY-II}}(\tilde{r}) &= \frac{-2 \ell (\ell+1)+2 s^2+1}{2 \tilde{r}^4}  -\frac{-2 \ell (\ell+1)+s (2 s+3)+5}{2 \tilde{r}^5} + \frac{6 s+9}{4 \tilde{r}^6}, \label{model-II-U1}\\
\nonumber U_{\textrm{ZLMY-II}}^{(2)}(\tilde{r}) & =  \frac{4 \ell (\ell+1)-4 s^2-5}{4 \tilde{r}^6} + \frac{-4 \ell (\ell+1)+s(4s+3)+12}{2 \tilde{r}^7} \\ & + \frac{16 \ell (\ell+1)-16 s (s+3)-139}{16 \tilde{r}^8} +  \frac{3 (8 s+21)}{16 \tilde{r}^9},\\
\nonumber  U_{\textrm{ZLMY-II}}^{(3)}(x) & =  \frac{- \ell (\ell+1)+ s^2+2}{ \tilde{r}^8}  + \frac{6 \ell (\ell+1)-3 s (2 s+1)-22}{2\tilde{r}^9}   \\ & + \frac{-24 \ell (\ell+1)+12 s (2 s+3)+173}{8 \tilde{r}^{10}} + \frac{4 \ell (\ell+1)-2 s (2 s+9)-73}{4 \tilde{r}^{11}} + \frac{3 (4 s+15)}{8 \tilde{r}^{12}}.
\end{align}
As it is obvious, the series expansion of the effective potential for the ZLMY-II model is very similar to the series expansion of the effective potential for the ZLMY-I. Therefore, we expect quite similar behavior for the TLNs studying the ZLMY-II model. Having the effective potential encoding quantum gravitational effects, we proceed to compute the TLNs for the ZLMY-II model using Green's function techniques used in the previous section.

\subsubsection{Scalar field response}

Here, we compute the TLNs in response to a scalar field perturbation for the ZLMY-II model. Having the first order effective potential (\ref{model-II-U1}) and the zeroth order solutions (\ref{solution-s=0-l=0-1}) and (\ref{solution-s=0-l=0-2}) for $\ell=0$, the first order correction to the TLN in response to a scalar field is obtained from the following integration
\begin{align}
    I[\mathcal{S}^{(1)}(\tilde{r})] = - \int_{1}^{\tilde{r}} \Phi^{(0)}_{+}(\tilde{r}^{\prime}) \mathcal{S}^{(1)}(\tilde{r}^{\prime}) \mathrm{d} \tilde{r}^{\prime} = \frac{3}{4 \tilde{r}^3}-\frac{5}{4 \tilde{r}^2}+\frac{1}{2 \tilde{r}}.
\end{align}
From which one observes that it has no constant term, indicating that the TLN at the first order, $\mathcal{O}(\tilde{\xi}^2)$, is vanishing, i.e., $\kappa_{0}^{0(1)} = 0$. This is similar to the ZLMY-I and AOS models predicting vanishing TLNs in this case. Extending the results up to the third order, $\mathcal{O}(\tilde{\xi}^6)$, we find that the TLN is vanishing, i.e., $\kappa_{0}^{0}=0$. 

For $\ell=1$, the zeroth order growing and decaying modes are given by (\ref{solution-s=0-l=1-1}) and (\ref{solution-s=0-l=1-2}), and the first order correction to the TLN can be extracted from the following integration
\begin{align}
    I[\mathcal{S}^{(1)}(\tilde{r})] = -\frac{1}{3} \int_{1}^{\tilde{r}} \Phi^{(0)}_{+}(\tilde{r}^{\prime}) \mathcal{S}^{(1)}(\tilde{r}^{\prime}) \mathrm{d} \tilde{r}^{\prime} = \frac{1}{16 \tilde{r}^3}-\frac{19}{48 \tilde{r}^2}+\frac{19}{24 \tilde{r}} -\frac{23}{24} -\frac{\log (\tilde{r})}{3} +\frac{\tilde{r}}{2},
\end{align}
implying that the first order correction to the TLN is $\kappa_{1}^{0(1)} = -(23/24 + \log(\tilde{r})/3)$. This TLN is a negative at the first order, similar to the ZLMY-I model, while it has negative running in contrast to the ZLMY-I model, which has a positive running in this case. This is also different from the AOS model, which predicts no running. Computing the TLN up to the second order, $\mathcal{O}(\tilde{\xi}^4)$, one finds that
\begin{align}
    \kappa_{1}^{0} = \left[-\frac{23}{24}\tilde{\xi}^2 + \left(\frac{23 \pi ^2}{36}-\frac{3461}{2880}\right)\tilde{\xi}^4 + \mathcal{O}(\tilde{\xi}^6) \right] + \left[-\frac{1}{3}\tilde{\xi}^2 + \left(\frac{4}{9} \pi ^2 +2 \right) \tilde{\xi}^4 + \mathcal{O}(\tilde{\xi}^6) \right]\log\left(\tilde{r}\right),
\end{align}
from which one can see that the second order solution positively contributes to both the TLN and the running. 

Finally, for $\ell=2$, the zeroth order solutions are given by (\ref{solution-s=0-l=2-1}) and (\ref{solution-s=0-l=2-2}), and the first order correction to the TLN can be obtained from
\begin{align}
    \nonumber I[\mathcal{S}^{(1)}(\tilde{r})] &= -\frac{1}{5} \int_{1}^{\tilde{r}} \Phi^{(0)}_{+}(\tilde{r}^{\prime}) \mathcal{S}^{(1)}(\tilde{r}^{\prime}) \mathrm{d} \tilde{r}^{\prime} \\ & = \frac{1}{240 \tilde{r}^3}-\frac{47}{720 \tilde{r}^2}+\frac{121}{360 \tilde{r}}-\frac{193}{120}-\frac{2 \log (\tilde{r})}{5}+\frac{29 \tilde{r}}{12}-\frac{29 \tilde{r}^2}{20} + \frac{11 \tilde{r}^3}{30},
\end{align}
denoting that $\kappa_{2}^{0(1)} =  -(193/120+2\log(\tilde{r})/5)$. This TLN is also negative at the first order similar to $\ell=1$ and the ZLMY-I model. However, it has again a negative running in contrast to the ZLMY-I model. Computing the TLN up to the second order, we find 
\begin{align}
    \kappa_{2}^{0} = \left[-\frac{193}{120}\tilde{\xi}^2 +  \left(\frac{393103}{4800}+\frac{193 \pi ^2}{10}\right) \tilde{\xi}^4 + \mathcal{O}(\tilde{\xi}^6) \right] + \left[-\frac{2}{5}  \tilde{\xi}^2 + \left(\frac{652+24 \pi ^2}{5} \right) \tilde{\xi}^4 + \mathcal{O}(\tilde{\xi}^6) \right] \log(\tilde{r}),
\end{align}
which demonstrates that the second-order solution contributes positively to both the TLN and its running, similar to the case of $\ell=1$. Therefore, our results for the TLNs in response to the scalar field perturbation reveal that the TLN vanishes for $\ell=0$ up to the third order. However, the TLNs for the next two lowest multipole numbers, $\ell=1$ and $\ell=2$, are negative with positive running at the leading order.

\subsubsection{Vector field response}

Now we calculate the TLNs in response to a vector field perturbation, $s=1$. For $\ell=1$, the zeroth order solutions are given by (\ref{solution-s=1-l=1-1}) and (\ref{solution-s=1-l=1-2}), and the first order correction to the TLN can then be derived from the following integration
\begin{align}
    I[\mathcal{S}^{(1)}(\tilde{r})] = -\frac{1}{3} \int_{1}^{\tilde{r}} \Phi^{(0)}_{+}(\tilde{r}^{\prime}) \mathcal{S}^{(1)}(\tilde{r}^{\prime}) \mathrm{d} \tilde{r}^{\prime} =\frac{5}{4 \tilde{r}}-\frac{17}{12} +\log (\tilde{r})+\frac{\tilde{r}}{6}.
\end{align}
This indicates that the first order correction to the TLN is $\kappa_{1}^{1(1)} = -17/12 + \log(\tilde{r})$, which is negative and has a positive running. In this case, the TLN has a similar behavior to the ZLMY-I model. However, this is in contrast to the AOS model predicting no running for $s=1$ and $\ell=1$. Computing the TLN up to the second order, we have
\begin{align}
    \kappa_{1}^{1} = \left[-\frac{17}{12}\tilde{\xi}^2 - \left(\frac{185+34 \pi^2}{24}\right) \tilde{\xi}^4 + \mathcal{O}(\tilde{\xi}^6) \right] + \left[\tilde{\xi}^2 + \left(\frac{\pi ^2-1}{2}  \right)\tilde{\xi}^4 + \mathcal{O}(\tilde{\xi}^6) \right] \log\left(\tilde{r}\right),
\end{align}
from which one can see that the second order solution negatively effects the TLN, while positively contributing to the running. 

For $\ell=2$, the zeroth order growing and decaying modes are given by (\ref{solution-s=1-l=2-1}) and (\ref{solution-s=1-l=2-2}), and the first order TLN can be found from the following integration
\begin{align}
 I[\mathcal{S}^{(1)}(\tilde{r})] &= -\frac{1}{5} \int_{1}^{\tilde{r}} \Phi^{(0)}_{+}(\tilde{r}^{\prime}) \mathcal{S}^{(1)}(\tilde{r}^{\prime}) \mathrm{d} \tilde{r}^{\prime} = \frac{3 \tilde{r}^3}{10}-\frac{31 \tilde{r}^2}{40}+\frac{9 \tilde{r}}{160}+\frac{81 \log (\tilde{r})}{80}-\frac{1}{320}+\frac{27}{64 \tilde{r}}.
\end{align}
Hence, the first order correction to the TLN is $\kappa_{2}^{1(1)} = -1/320 + 81 \log(\tilde{r})/80$, which is negative in contrast to the ZLMY-I model and similar to the AOS model. However, it has a positive running at the first order similar to the ZLMY-I. Computing the TLN up to the second order, we obtain 
\begin{align}
    \kappa_{2}^{1} = \left[-\frac{1}{320} \tilde{\xi}^2 -\left(\frac{362987}{1920}+\frac{27 \pi ^2}{320}\right)\tilde{\xi}^4 + \mathcal{O}(\tilde{\xi}^6) \right] + \left[\frac{81 }{80} \tilde{\xi}^2 + \left(\frac{613}{32}+\frac{2187 \pi ^2}{80}\right)\tilde{\xi}^4 + \mathcal{O}(\tilde{\xi}^6) \right]\log\left(\tilde{r}\right),
\end{align}
implying that the second order solution negatively effects the TLN, while positively contributing to its running. 

Finally, given the zeroth order solutions in Eq. (\ref{solution-s=1-l=3-1}) and (\ref{solution-s=1-l=3-2}) for $\ell=3$, one can read the first order correction to the TLN from the following expression
\begin{align}
    \nonumber I[\mathcal{S}^{(1)}(\tilde{r})] &= -\frac{1}{7} \int_{1}^{\tilde{r}} \Phi^{(0)}_{+}(\tilde{r}^{\prime}) \mathcal{S}^{(1)}(\tilde{r}^{\prime}) \mathrm{d} \tilde{r}^{\prime} \\ & = \frac{3 \tilde{r}^5}{10}-\frac{5 \tilde{r}^4}{4}+\frac{2519 \tilde{r}^3}{1260}-\frac{433 \tilde{r}^2}{315}-\frac{13 \tilde{r}}{175}+\frac{72 \log (\tilde{r})}{175}+\frac{989}{3150}+\frac{3}{35 \tilde{r}}.
\end{align}
Hence, the first order correction to the TLN turns out to be $\kappa_{3}^{1(1)} = 989/3150 + 72 \log(\tilde{r})/175$, which is positive with a positive running. This is similar to the ZLMY-I black hole and in contrast to the AOS model predicting a negative TLN without logarithmic running. Computing the TLN up to the second order, one finds that
\begin{align}
   \nonumber  \kappa_{3}^{1} & = \left[\frac{989}{3150} \tilde{\xi}^2 + \left(\frac{5934 \pi ^2}{175}-\frac{71208721}{151200}\right)\tilde{\xi}^4 + \mathcal{O}(\tilde{\xi}^6) \right] \\ &+ \left[\frac{72 }{175}\tilde{\xi}^2  + \left(\frac{1431943}{6300}+\frac{7776 \pi ^2}{175}\right) \tilde{\xi}^4 + \mathcal{O}(\tilde{\xi}^6) \right]\log\left(
   \tilde{r}\right),
\end{align}
from which one can see that the second order solution positively contributes to both the TLN and the running. In summary, we find that the TLNs in response to the vector field perturbation for the ZLMY-II model are negative for the first two lowest multipole numbers, $\ell=1$ and $\ell=2$, and positive for the third lowest multipole number, $\ell=3$. However, the TLNs for all three lowest multipole numbers exhibit positive running at the leading order.

\begin{table}[t!]
\centering
\begin{tabular}{ c c c c c c c c c c}
\hline
\hline
 &   & $s=0$ &  &  & $s=1$ &  &  & $s=2$ &  \\ 
\hline
\ \ \ \ \ \  &  \ \ \  $\ell=0$  \ \ \  &  \ \ \  $\ell=1$  \ \ \  &  \ \ \  $\ell=2$  \ \ \  &  \ \ \  $\ell=1$  \ \ \ &  \ \ \  $\ell=2$  \ \ \  &  \ \ \  $\ell=3$  \ \ \  &  \ \ \  $\ell=2$  \ \ \  &  \ \ \ $\ell=3$  \ \ \  &  \ \ \  $\ell=4$  \ \ \  \\ 
\hline
$C(s, \ell)$ & $0$ & $-1.2385$ & $-2.0785$ & $-1.8308$ & $-0.0040$ & $ 0.4058$ & $0.7754$ & $0.1872$ & $0.0356$  \\ \hline
$C_{\textrm{log}}(s, \ell)$ &$0$ & $-0.4308$ & $-0.5169$ & $1.2923$ & $1.3085$ & $0.5317$ & $0$ & $0$ & $0$ \\
\hline
\hline
\end{tabular}
\caption{The leading order corrections to the TLNs and their running, i.e., Eq. (\ref{TLN-model-I-II}), for the ZLMY-II loop quantum black hole model and the three lowest multipole numbers $\ell$ in response to the scalar, vector, and axial gravitational field perturbations.  }
\label{TLN-first-order-model-II}
\end{table}

\subsubsection{Axial gravitational field response}

In this section, we compute the TLNs for the axial gravitational field response for the ZLMY-II model. For $\ell=2$, the zeroth order solutions are given by Eqs. (\ref{solution-s=2-l=2-1}) and  (\ref{solution-s=2-l=2-2}), and we can compute the following integration
\begin{align}
     I[\mathcal{S}^{(1)}(\tilde{r})] &= -\frac{1}{5} \int_{1}^{\tilde{r}} \Phi^{(0)}_{+}(\tilde{r}^{\prime}) \mathcal{S}^{(1)}(\tilde{r}^{\prime}) \mathrm{d} \tilde{r}^{\prime} = \frac{\tilde{r}^3}{10}+\frac{7 \tilde{r}^2}{20}-\frac{21 \tilde{r}}{20}+\frac{3}{5}.
\end{align}
This indicates that the first order correction to the TLN is $\kappa_{2}^{2(1)}=3/5$, which is positive and has no logarithmic running similar to the ZLMY-I model. However, this is in contrast to the AOS model, which predicts a negative TLN. Extracting the TLN up to the third order, we find that
\begin{align}
    \kappa_{2}^{2} = \left[\frac{3}{5}  \tilde{\xi}^2 +  \frac{25}{16} \tilde{\xi}^4  - \left(\frac{3419}{1200}+\frac{3 \pi ^2}{10} \right)\tilde{\xi}^6 + \mathcal{O}(\tilde{\xi}^8) \right]  + \left[-\frac{3}{5} \tilde{\xi}^4 -\frac{101}{40} \tilde{\xi}^6 + \mathcal{O}(\tilde{\xi}^8)  \right] \log (\tilde{r}),
\end{align}
in which one can observe that the second order solution positively effects the TLN, while negatively contributing to the running.  

For $\ell=3$, the zeroth order solutions are given by Eqs. (\ref{solution-s=1-l=3-1}) and (\ref{solution-s=1-l=3-2}), and the first order correction to the TLN can be extracted from the constant part of the following integration 
\begin{align}
     I[\mathcal{S}^{(1)}(\tilde{r})] &= -\frac{1}{7} \int_{1}^{\tilde{r}} \Phi^{(0)}_{+}(\tilde{r}^{\prime}) \mathcal{S}^{(1)}(\tilde{r}^{\prime}) \mathrm{d} \tilde{r}^{\prime} = \frac{3 \tilde{r}^5}{14}-\frac{15 \tilde{r}^4}{28}+\frac{11 \tilde{r}^3}{56}+\frac{505 \tilde{r}^2}{1008}-\frac{25 \tilde{r}}{48}+\frac{73}{504}.
\end{align}
This implies that the first order correction to the TLN is $\kappa_{3}^{2} = 73/504$, which is again positive and has no running at the leading order similar to $\ell=2$. Computing the TLN up to the third order, one finds that
\begin{align}
\nonumber \kappa_{3}^{2} & = \left[\frac{73}{504} \tilde{\xi}^2 +  \frac{35341}{15120} \tilde{\xi}^4 - \left(\frac{213278731}{7257600}+\frac{2701 \pi ^2}{1512}\right) \tilde{\xi}^6  + \mathcal{O}(\tilde{\xi}^8) \right] \\ & + \left[-\frac{37}{126} \tilde{\xi}^4 -\frac{111271}{5040} \tilde{\xi}^6 + \mathcal{O}(\tilde{\xi}^8) \right] \log (\tilde{r}),
\end{align}
from which one again finds that the second order solution positively contributes to the TLN and has a negative impact on the running, similar to $\ell=2$ and the ZLMY-I model. 

For $\ell=4$, having the zeroth order solutions (\ref{solution-s=2-l=4-1}) and (\ref{solution-s=2-l=4-2}), one can compute the following integration
\begin{align}
\nonumber I[\mathcal{S}^{(1)}(\tilde{r})] &= -\frac{1}{9} \int_{1}^{\tilde{r}} \Phi^{(0)}_{+}(\tilde{r}^{\prime}) \mathcal{S}^{(1)}(\tilde{r}^{\prime}) \mathrm{d} \tilde{r}^{\prime}\\ & = \frac{31 \tilde{r}^7}{126}-\frac{19 \tilde{r}^6}{18}+\frac{1451 \tilde{r}^5}{840}-\frac{137 \tilde{r}^4}{112}+\frac{677 \tilde{r}^3}{4704}+\frac{135 \tilde{r}^2}{448}-\frac{75 \tilde{r}}{448}+\frac{647}{23520},
\end{align}
implying that $\kappa_{4}^{2(1)} =  647/23520$, which is positive and has no running similar to $\ell=2$ and $\ell=3$. Computing the TLN up to the third order, we have
\begin{align}
   \nonumber  \kappa_{4}^{2} & = \left[\frac{647}{23520} \tilde{\xi}^2 + \frac{159629}{70560} \tilde{\xi}^4 + \left(\frac{2910854179}{37632000}+\frac{322853 \pi ^2}{49000}\right)\tilde{\xi}^6 + \mathcal{O}(\tilde{\xi}^8)  \right] \\ & + \left[\frac{499}{4900} \tilde{\xi}^4 + \frac{146245433}{2940000} \tilde{\xi}^6 + \mathcal{O}(\tilde{\xi}^8) \right] \log (\tilde{r}),
\end{align}
in which the second and third order solutions positively contribute to both the TLN and the running. Hence, we found that the TLNs in response to axial gravitational field perturbation are positive for the first three multipole numbers and has no logarithmic running at the leading order. However, they exhibit either a positive or negative running at the second order depending on the multipole number.

Finally, we should mention that the first order corrections to the TLNs in a compact form is also given by Eq. (\ref{TLN-model-I-II}) for the ZLMY-II model. We again find that $\alpha=2$ indicating that the TLNs are also Planck scale suppressed with a mass scaling similar to the ZLMY-I model. We record the values for $C(s, \ell)$ and $C_{\textrm{log}}(s, \ell)$ in Table \ref{TLN-first-order-model-II} for comparison. From Table \ref{TLN-first-order-model-II}, one can find that considering the same black hole mass, the largest magnitude for the TLNs is obtained for scalar field response with $\ell=2$. On the other hand, the TLN in response to the vector field response with $\ell=2$ has the smallest magnitude. Moreover, the largest magnitude for the running is for the vector TLN with $\ell=2$.

\subsection{ABV model}\label{Section IV-C}

In this section, we extract the TLNs for the ABV loop quantum black hole model \cite{Alonso-Bardaji:2021yls} in response to a scalar field, a vector field, and an axial gravitational field response. We Write the metric functions in terms of dimensionless quantities as follows
\begin{align}\label{HC-metirc}
    A(\tilde{r}) = f(\tilde{r}) = 1- \frac{1}{{\tilde{r}}}, \ \ \ \  \ \ \ \ \ B(\tilde{r}) = \left(1- \frac{\tilde{r}_{0}}{\tilde{r}} \right)f(\tilde{r}),
\end{align}
where $\tilde{r}_{0}=r_{0}/r_{H} = \lambda^2/(1+\lambda^2) = \sqrt{\frac{\Delta}{16\pi}}
M^{-1}$ and $\Delta$ is the minimum eigenvalue of the area operator given in Section \ref{Section II}. The parameter $\tilde{r}_{0}$ takes small values for astrophysical black holes, i.e., $\tilde{r}_{0}\ll 1$ or equivalently $\lambda^2 \ll 1$ \cite{Alonso-Bardaji:2021yls}. One can then transform Eq.(\ref{perturbation-r}), in the static limit, into Eq. (\ref{perturbation-transformed}) by the following field redefinition
\begin{align}
    \Psi(\tilde{r}) =  \frac{\Phi(\tilde{r})}{\sqrt{Z(\tilde{r})}},  \ \  \ \ \ \ \ \ \ \ \ Z(r) = \sqrt{1- \frac{\tilde{r}_{0}}{\tilde{r}}}.
\end{align}
Then, the zeroth order effective potential $U^{(0)}(\tilde{r})$ is given by Eq. (\ref{zero-order-potential}). The first, second, and third order effective potentials are given by 
\begin{align}
U^{(1)}_{\textrm{ABV}}(\tilde{r}) &= \frac{2 \ell(\ell+1) + s(1-2s)}{2 \tilde{r}^3} + \frac{2 s+1}{4 \tilde{r}^4}, \label{hc-U1}\\
U_{\textrm{ABV}}^{(2)}(\tilde{r}) &=  \frac{16 \ell(\ell+1)+8 s (1-2 s)-3}{16\tilde{r}^4} + \frac{8s+7}{16 \tilde{r}^5},\\
U_{\textrm{ABV}}^{(3)}(\tilde{r}) &=  \frac{8 \ell(\ell+1)+4 s (1-2 s)-3}{8\tilde{r}^5} + \frac{4s+5}{8\tilde{r}^6}.
\end{align}
This implies that the effective potential for the ABV model has a completely different series expansion in comparison with ZLMY-I/II models. As an example, the first order effective potential has corrections from $\mathcal{O}(\tilde{r}^{-3})$ up to $\mathcal{O}(\tilde{r}^{-4})$, while the first order effective potential for the ZLMY-I/II models has corrections from $\mathcal{O}(\tilde{r}^{-4})$ up to $\mathcal{O}(\tilde{r}^{-6})$. It is also interesting that the first order effective potential has a correction of $\mathcal{O}(\tilde{r}^{-3})$, which also appears in the zeroth order effective potential. Due to such different behavior for the effective potential, one expects that the ABV model will exhibit distinct behavior in their TLNs. Given the perturbative modifications to the effective potential and zeroth order solutions (\ref{zero-order-solutions-1}) and (\ref{zero-order-solutions-2}), one can find the higher order solutions and then extract the corresponding TLNs for all three scalar, vector, and axial gravitational field perturbations, as we have already carried out that analysis for the ZLMY-I/II in the previous sections.

\subsubsection{Scalar field response}

Here, we compute the TLNs in response to the scalar field perturbation for the ABV model. For $s=0$ and $\ell=0$, one can compute the first order correction to the TLN from the following integration
\begin{align}
    I[\mathcal{S}^{(1)}(\tilde{r})] = - \int_{1}^{\tilde{r}} \Phi^{(0)}_{+}(\tilde{r}^{\prime}) \mathcal{S}^{(1)}(\tilde{r}^{\prime}) \mathrm{d} \tilde{r}^{\prime} = -\frac{1}{4} + \frac{1}{4 \tilde{r}},
\end{align}
implying that $\kappa_{0}^{0(1)} = -1/4$. This TLN is negative and has no logarithmic running. This is in contrast to the ZLMY-I/II and the AOS models predicting vanishing TLNs for $\ell=0$. Computing the TLN up to the third order, $\tilde{r}_{0}^3$, one finds that
\begin{align}
    \kappa_{0}^{0} = - \frac{1}{4}\tilde{r}_{0} + \mathcal{O}(\tilde{r}_{0}^4).
\end{align}
From this, one can see that the TLN is negative up to the third order, and there is no logarithmic running.

For $\ell=1$, the zeroth order growing and decaying modes are given by Eqs. (\ref{solution-s=0-l=1-1}) and (\ref{solution-s=0-l=1-2}), and one can read the first order correction to the TLN from the following expression
\begin{align}
    I[\mathcal{S}^{(1)}(\tilde{r})] = -\frac{1}{3} \int_{1}^{\tilde{r}} \Phi^{(0)}_{+}(\tilde{r}^{\prime}) \mathcal{S}^{(1)}(\tilde{r}^{\prime}) \mathrm{d} \tilde{r}^{\prime} = -\frac{\tilde{r}^2}{3}+\frac{7 \tilde{r}}{12}  -\frac{\log (\tilde{r})}{12} -\frac{13}{48} +\frac{1}{48 \tilde{r}},
\end{align}
which indicates $\kappa_{1}^{0(1)} = -13/48 - \log(\tilde{r})/12$. This means the TLN is negative with negative running similar to the ZLMY-II model and different from the ZLMY-I model predicting a positive running. Moreover, the presence of running behavior is in contrast to the AOS model with no running for $\ell=1$. Computing the TLN up to the second order, one finds that
\begin{align}\label{TLN-s=0-l=1}
    \kappa_{1}^{0} =  \left[-\frac{13}{48}\tilde{r}_{0} +  \left(\frac{11}{32}+\frac{13 \pi ^2}{144}\right)\tilde{r}_{0}^2 + \mathcal{O}(\tilde{r}_{0}^3) \right] + \left[-\frac{1}{12}\tilde{r}_{0}+ \left(\frac{3}{8}+\frac{\pi ^2}{36}\right)  \tilde{r}_{0}^2 + \mathcal{O}(\tilde{r}_{0}^3) \right]\log \left(\tilde{r}\right),
\end{align}
implying that the second order solution positively contributes to both the TLN and the running.

Finally, the zeroth order solutions for $\ell=2$ are given by Eqs. (\ref{solution-s=0-l=2-1}) and (\ref{solution-s=0-l=2-2}), and the first order correction to the TLN can be extracted from the following integration
\begin{align}
    I[\mathcal{S}^{(1)}(\tilde{r})] = -\frac{1}{5} \int_{1}^{\tilde{r}} \Phi^{(0)}_{+}(\tilde{r}^{\prime}) \mathcal{S}^{(1)}(\tilde{r}^{\prime}) \mathrm{d} \tilde{r}^{\prime} =-\frac{3 \tilde{r}^4}{10}+\frac{47 \tilde{r}^3}{60}-\frac{3 \tilde{r}^2}{4}+\frac{\tilde{r}}{3}-\frac{\log (\tilde{r})}{60}-\frac{49}{720} +\frac{1}{720 \tilde{r}},
\end{align}
denoting that $\kappa_{2}^{0(1)} = -49/720-\log(\tilde{r})/60$. This TLN is negative and has a negative running similar to $\ell=2$ and the ZLMY-II model. Computing the second order correction to the TLN, one finds the TLN up to the second order as follows
\begin{align}
    \kappa_{2}^{0} = \left[-\frac{49}{720} \tilde{r}_{0} + \left(\frac{1243+196\pi^2}{2880}\right)\tilde{r}_{0}^2 + \mathcal{O}(\tilde{r}_{0}^3) \right] + \left[-\frac{1}{60} \tilde{r}_{0}  +\left(\frac{22+\pi^2}{30}\right)\tilde{r}_{0}^2 + \mathcal{O}(\tilde{r}_{0}^3) \right]\log \left(\tilde{r}\right) .
\end{align}
This 
demonstrates that the second order solution positively contributes to both the TLN and its running, similar to $\ell=2$. Therefore, our results for the scalar field response reveal that the scalar TLNs are negative for the first lowest multipole numbers with a negative running at the leading order, sharing similarity with ZLMY-II model in the case of scalar field response. 

\subsubsection{Vector field response}

Now, we compute the TLNs in response to the vector field perturbation for the ABV model. Having the zeroth order solutions (\ref{solution-s=1-l=1-1}) and (\ref{solution-s=1-l=1-2}) for $\ell=1$, the first order correction to the TLN is given by the constant part of the following integration
\begin{align}
    I[\mathcal{S}^{(1)}(\tilde{r})] = -\frac{1}{3} \int_{1}^{\tilde{r}} \Phi^{(0)}_{+}(\tilde{r}^{\prime}) \mathcal{S}^{(1)}(\tilde{r}^{\prime}) \mathrm{d} \tilde{r}^{\prime} = -\frac{\tilde{r}^2}{4}-\frac{\tilde{r}}{4}+\frac{1}{2},
\end{align}
which turns out to be $\kappa_{1}^{1(1)} = 1/2$. This TLN is positive and has no running at the leading order. This is in contrast to the ZLMY-I/II and the AOS models, which have negative TLNs for $\ell=1$ with either a positive, negative, or vanishing running. Computing the TLN up to the third order, one finds that 

\begin{align}
\kappa_{1}^{1}& = \left[\frac{1}{2} \tilde{r}_{0} + \frac{1}{4} \tilde{r}_{0}^2 + \left(\frac{21}{128}+\frac{\pi ^2}{16}\right)\tilde{r}_{0}^3 + \mathcal{O}(\tilde{r}_{0}^4) \right] + \left[\frac{1 }{4}\tilde{r}_{0}^2 + \frac{9 }{32} \tilde{r}_{0}^3 + \mathcal{O}(\tilde{r}_{0}^4) \right]\log \left(\tilde{r}\right).
\end{align}
This result shows that the TLN for $s=1$ and $\ell=1$ is positive up to the third order and exhibits a positive logarithmic running at the second order. 

For $\ell=2$, the zeroth order growing and decaying modes are given by (\ref{solution-s=1-l=2-1}) and (\ref{solution-s=1-l=2-2}), leading to the following integration for the first order correction to the TLN
\begin{align}
    I[\mathcal{S}^{(1)}(\tilde{r})] = -\frac{1}{5} \int_{1}^{\tilde{r}} \Phi^{+}(\tilde{r}^{\prime}) \mathcal{S}^{(1)}(\tilde{r}^{\prime}) \mathrm{d} \tilde{r}^{\prime} =-\frac{11 \tilde{r}^4}{40}+\frac{\tilde{r}^3}{2}-\frac{63 \tilde{r}^2}{320}-\frac{27 \tilde{r}}{320}+\frac{9}{160}.
\end{align}
Thus, we find  that $\kappa_{1}^{2(1)} = 9/160$, which is positive and has no running similar to $\ell=1$. Calculating the TLN up to the third order, one obtains
\begin{align}
\kappa_{2}^{1} &= \left[\frac{9}{160}\tilde{r}_{0} + \frac{153}{640} \tilde{r}_{0}^2 + \left(\frac{6467}{10240}+\frac{81 \pi ^2}{1280}\right) \tilde{r}_{0}^3 + \mathcal{O}(\tilde{r}_{0}^4) \right] + \left[ \frac{27}{320} \tilde{r}_{0}^2 + \frac{351}{512} \tilde{r}_{0}^{3} + \mathcal{O}(\tilde{r}_{0}^4) \right]\log \left(\tilde{r}\right).
\end{align}
This shows that the second and third order solutions positively contribute to both the TLN and the running. 

Finally, for $\ell=3$, the the first order TLN can be extracted from the following expression
\begin{align}
    I[\mathcal{S}^{(1)}(\tilde{r})] = -\frac{1}{7} \int_{1}^{\tilde{r}} \Phi^{(0)}_{+}(\tilde{r}^{\prime}) \mathcal{S}^{(1)}(\tilde{r}^{\prime}) \mathrm{d} \tilde{r}^{\prime} =-\frac{23 \tilde{r}^6}{84}+\frac{359 \tilde{r}^5}{420}-\frac{311 \tilde{r}^4}{315}+\frac{31 \tilde{r}^3}{63}-\frac{13 \tilde{r}^2}{175}-\frac{3 \tilde{r}}{175}+\frac{1}{175},
\end{align}
denoting that $\kappa_{3}^{1(1)} = 1/175$. This TLN is also positive with no running similar to $\ell=1$ and $\ell=2$. One can then calculate the TLN up to the third order as follows
\begin{align}
\kappa_{3}^{1} &= \left[\frac{1}{175}\tilde{r}_{0} + \frac{93}{1400} \tilde{r}_{0}^2 + \left(\frac{51203}{100800}+\frac{9 \pi ^2}{350}\right)\tilde{r}_{0}^3 + \mathcal{O}(\tilde{r}_{0}^4) \right] + \left[\frac{3 }{175}\tilde{r}_{0}^2 + \frac{549}{1400} \tilde{r}_{0}^3 + \mathcal{O}(\tilde{r}_{0}^4)  \right]\log \left(\tilde{r}\right).
\end{align}
indicating that the second and third order solutions positively contribute to both the TLN and the running. Therefore, we found that the vector TLNs for the three lowest multipole numbers are positive and have no running behavior at the leading order in the case of the ABV model. However, they exhibit positive running at the second order.

\begin{table}[t!]
\centering
\begin{tabular}{ c c c c c c c c c c}
\hline
\hline
 &   & $s=0$ &  &  & $s=1$ &  &  & $s=2$ &  \\ 
\hline
\ \ \ \ \ \  &  \ \ \  $\ell=0$  \ \ \  &  \ \ \  $\ell=1$  \ \ \  &  \ \ \  $\ell=2$  \ \ \  &  \ \ \  $\ell=1$  \ \ \ &  \ \ \  $\ell=2$  \ \ \  &  \ \ \  $\ell=3$  \ \ \  &  \ \ \  $\ell=2$  \ \ \  &  \ \ \ $\ell=3$  \ \ \  &  \ \ \  $\ell=4$  \ \ \  \\ 
\hline
$C(s, \ell)$ & $-0.0802$ & $-0.0868$ & $-0.0218$ & $0.1603$ & $0.0181$ & $0.0018$ & $0.0748$ & $0.0037$ & $0.0002$  \\ \hline
$C_{\textrm{log}}(s, \ell)$ &$0$ & $-0.0267$ & $-0.0054$ & $0$ & $0$ & $0$ & $0$ & $0$ & $0$ \\
\hline
\hline
\end{tabular}
\caption{The leading order corrections to the TLNs and their running, i.e., Eq. (\ref{TLN-model-I-II}), for the ABV loop quantum black hole model and the three lowest multipole numbers $\ell$ in response to the scalar, vector, and axial gravitational field perturbations. }
\label{TLN-first-order-hc}
\end{table}

\subsubsection{Axial gravitational field response}

In this section, we calculate the TLNs for axial gravitational field perturbation in the case of the ABV model. For $s=2$ and $\ell=2$, the zeroth order solutions are given by Eqs. (\ref{solution-s=2-l=2-1}) and (\ref{solution-s=2-l=2-2}) and the first order correction to the TLN is obtained from the constant part of the following expression
\begin{align}
    I[\mathcal{S}^{(1)}(\tilde{r})] = -\frac{1}{5} \int_{1}^{\tilde{r}} \Phi^{(0)}_{+}(\tilde{r}^{\prime}) \mathcal{S}^{(1)}(\tilde{r}^{\prime}) \mathrm{d} \tilde{r}^{\prime} =-\frac{3 \tilde{r}^4}{20}-\frac{\tilde{r}^3}{12}+\frac{7}{30}.
\end{align}
Hence, the TLN, for $\ell=2$ at the first order, is $\kappa_{2}^{2(1)} = 7/30$. This TLN is positive and has no logarithmic running behavior at the first order, similar to the ZLMY-I/II models, and in contrast to the AOS model with a negative TLN. Computing the TLN up to the third order, one finds that
\begin{align}
\kappa_{2}^{2} &= \left[\frac{7}{30}\tilde{r}_{0} -\frac{63}{160}\tilde{r}_{0}^2 + \left(\frac{343 \pi ^2}{6480}-\frac{4021}{6912}+\right) \tilde{r}_{0}^3  
 + \mathcal{O}(\tilde{r}_{0}^4) \right] + \left[\frac{49 }{180} \tilde{r}_{0}^{2} + \frac{2303}{25920} \tilde{r}_{0}^{3} + \mathcal{O}(\tilde{r}_{0}^4) \right]\log \left(\tilde{r}\right),
 \end{align}
which demonstrates that the second and third order solutions negatively effects the TLN, while positively contributing to the running. 

For $\ell=3$, the first order correction to the TLN can be extracted from the constant part of the following integration
\begin{align}
    I[\mathcal{S}^{(1)}(\tilde{r})] = -\frac{1}{7} \int_{1}^{\tilde{r}} \Phi^{(0)}_{+}(\tilde{r}^{\prime}) \mathcal{S}^{(1)}(\tilde{r}^{\prime}) \mathrm{d} \tilde{r}^{\prime} =-\frac{3 \tilde{r}^6}{14}+\frac{11 \tilde{r}^5}{28}-\frac{25 \tilde{r}^4}{168}-\frac{125 \tilde{r}^3}{3024}+\frac{5}{432},
\end{align}
which turns out to be $\kappa_{3}^{2(1)} = 5/432$. This TLN is positive and has no running at the leading order similar to $\ell=2$ and also the ZLMY-I/II models and in contrast to the AOS model predicting a negative TLN. Computing the TLN up to the third order, we find that
\begin{align}
\kappa_{3}^{2} &= \left[\frac{5}{432} \tilde{r}_{0} -\frac{47}{6912} \tilde{r}_{0}^2 + \left(\frac{6125 \pi ^2}{373248}-\frac{331361}{1658880} \right)\tilde{r}_{0}^3 + \mathcal{O}(\tilde{r}_{0}^4) \right] + \left[\frac{175}{5184} \tilde{r}_{0}^2 + \frac{39025}{373248} \tilde{r}_{0}^3 + \mathcal{O}(\tilde{r}_{0}^4) \right]\log \left(\tilde{r}\right),
\end{align}
indicating that the second and third order solutions negatively impact TLN, while positively contributing to the running. 

Finally, for $\ell=4$, the first order TLN can be computed as follows
\begin{align}
    \nonumber I[\mathcal{S}^{(1)}(\tilde{r})] &= -\frac{1}{9} \int_{1}^{\tilde{r}} \Phi^{(0)}_{+}(\tilde{r}^{\prime}) \mathcal{S}^{(1)}(\tilde{r}^{\prime}) \mathrm{d} \tilde{r}^{\prime} \\ & =-\frac{17 \tilde{r}^8}{72}+\frac{199 \tilde{r}^7}{252}-\frac{41 \tilde{r}^6}{42}+\frac{173 \tilde{r}^5}{336}-\frac{125 \tilde{r}^4}{1568}-\frac{125 \tilde{r}^3}{9408}+\frac{1}{1344},
\end{align}
implying that $\kappa_{4}^{2(1)} = 1/1344$. This TLN is positive and does not exhibit a logarithmic running at the leading order similar to $\ell=2$ and $\ell=3$ and also the ZLMY-I/II models. Calculating the TLN up to the third order, we obtain 
\begin{align}
\kappa_{4}^{2} &= \left[\frac{1}{1344} \tilde{r}_{0} + \frac{79}{35840} \tilde{r}_{0} ^2 +  \left(\frac{7 \pi ^2}{1024}-\frac{5419391}{180633600}\right)\tilde{r}_{0} ^{3} + \mathcal{O}(\tilde{r}_{0}^4) \right] + \left[\frac{1 }{256} \tilde{r}_{0} ^2 + \frac{71}{2048} \tilde{r}_{0} ^{3} + \mathcal{O}(\tilde{r}_{0}^4)\right] \log \left(\tilde{r}\right),
\end{align}
from which we can see that the second and third order solutions positively contribute to both the TLN and the running. Hence, we observed that the tensor TLNs are positive and do not exhibit running behavior at the leading order for the ABV model. 

We can simply write the first order corrections to the TLNs for the ABV model as given by Eq. (\ref{TLN-model-I-II}) with $\alpha=1$. This means that the magnitude of the TLNs is Planck scale suppressed, but it has a larger magnitude for the same black hole mass in comparison with ZLMY-I/II models. The values of $C(s, \ell)$ and $C_{\textrm{log}}(s, \ell)$ are recorded in Table \ref{TLN-first-order-hc} for comparison. From Table \ref{TLN-first-order-hc}, we can see that the largest TLN is obtained for vector field response with $\ell=1$, while the smallest one occurs for the axial gravitational field with $\ell=4$. Moreover, the scalar TLN with $\ell=1$ has the largest magnitude for the running. \\

Let us conclude this section, by summarizing the key findings of our analysis for ZLMY-I/II and ABV models. Our findings, together with those in Ref. \cite{Motaharfar:2025typ}, confirm that the TLNs are, in general, nonzero for loop quantum black holes. This implies that loop quantum black holes are tidally deformed under the influence of an external tidal field and the TLNs carry information about the internal structure of loop quantum black holes.

The magnitude of the TLNs for all three models considered in this manuscript is Planck scale suppressed, i.e., $\kappa_{\ell}^{s}\propto M^{-2}$ (in Planck units) for the ZLMY-I/II models and $\kappa_{\ell}^{s}\propto M^{-1}$ for the ABV model, similar to the AOS model. Although in this case the suppression is stronger than in the AOS model, i.e., $\kappa_{\ell}^{s} \propto M^{-2/3}$. Such Planck scale suppression, with similar mass scaling observed for the ZLMY-I/II models, was also found in Ref. \cite{Kim:2020dif}, where quantum corrections to the TLNs for Schwarzschild black holes are obtained from a sum rule for the TLNs derived from quantum field theory computations at a semiclassical level. Therefore, one concludes that the TLNs are generally Planck scale suppressed for loop quantum black holes derived from effective dynamics. To sharpen this conclusion, we present the following arguments. First, based on physical intuition, the only dimensionless quantity that one can build with the only available parameter of the model, the mass of the black hole, and the fundamental scale of quantum gravity $M_{\textrm{pl}}$, is $M/{M_{\textrm{Pl}}}$. One then expects that the larger the black hole, the more difficult it is to perturb and eventually deform it; hence, the magnitude of the TLNs should be Planck scale suppressed, as is the case for loop quantum black holes. Second, quantum gravitational effects become important when the energy density reaches the Planck density, or equivalently for $r<(M/M_{\textrm{Pl}})^{1/3}l_{\textrm{Pl}}$. This implies that quantum gravitational effects are larger at the event horizon for smaller black holes, i.e., $ 2GM/c^2<(M/M_{\textrm{Pl}})^{1/3}l_{\textrm{Pl}}$ or equivalently, $(M_{\textrm{Pl}}/M)^{2/3}\gtrsim 1$. This suggests that the quantum gravitational corrections to the TLNs should be Planck scale suppressed at low curvature regime (event horizon) and of order $(M_{\textrm{Pl}}/M)^{2/3}$, similar to what was found for the AOS model. Therefore, all of these results and arguments support that the magnitude of the TLNs for loop quantum black holes should be Planck scale suppressed. This implies that loop quantum corrections to the TLNs are very small for astrophysical black holes and not detectable in near future.

Finally, our results also demonstrate that the sign and running behavior of the TLNs for loop quantum black holes serve as distinctive features of quantum ambiguities arising from loop quantization. In fact, extracted TLNs for loop quantum black holes can be either positive, negative, or zero, with their running exhibiting similar variability depending on the spin of the external tidal field, the multipole number, and the specific loop quantum black hole. This means that loop quantum black holes with different quantum ambiguities distinctively respond to an external tidal field, reflecting that their internal structures are distinct. Furthermore, the TLNs can also serve as a tool to distinguish loop quantum black holes from those predicted by modified gravity theories \cite{Katagiri:2024fpn} and by effective field theory of gravity \cite{Cardoso:2018ptl, Barura:2024uog}, as each framework predicts a distinct TLN behavior.

\section{Summary and conclusions}\label{Section V}

Classical GR predicts that black holes are the most resilient objects in the universe with precisely vanishing TLNs. This implies that classical black holes do not acquire induced moments in response to an external static tidal field. More precisely, classical black holes lack a distinct internal structure, strengthening the notion that they are bald. On the other hand, the existence of singularities in classical black holes, where classical GR breaks down, necessitates a theory of quantum gravity to resolve classical singularities and make physical predictions at the Planck scale. The resolution of interior singularities of classical black holes due to the discreteness of spacetime geometry suggests that the internal structure of quantum black holes is fundamentally distinct from that of classical counterparts. Hence, a pertinent question arises: Are loop quantum black holes tidally deformed under the influence of an external tidal field? 

This question was partially addressed in Ref. \cite{Motaharfar:2025typ}, in which it was observed that the TLNs for loop quantum black holes, specifically the AOS model, are generally nonzero, negative, and Planck scale suppressed. However, the AOS model is a homogeneous minisuperspace model primarily constructed for the quantization of macroscopic black holes. To understand the genericness of the nonzero TLNs for loop quantum black holes, we studied the linear static response of three covariant loop quantum black holes, the ZLMY-I/II and the ABV models, which are interestingly built to maintain general covariance independent of the gauge choice. Due to the complicated functional form of the metric functions, obtaining the exact solution for the radial component of the perturbations is not possible for these three loop quantum black holes as it was done for the AOS model. However, since the relation between the event horizon and the black hole mass is the same as for the classical Schwarzschild black hole, we were able to transform the equation for the radial component of perturbations in such a way that all the quantum gravitational effects are captured in the effective potential. Given the smallness of quantum parameters, we expanded the effective potential and perturbatively solved the equation for the radial component of perturbations by applying the Green's function technique developed in the literature \cite{Barura:2024uog}. Having the perturbative solutions, we extracted the scalar, vector, and tensor TLNs for all three loop quantum black holes and the three lowest multipole numbers up to the third order expansion in quantum parameters.

For the ZLMY-I loop quantum black hole model, we found that the TLNs are generally nonzero and Planck scale suppressed, i.e., $\kappa_{\ell}^{s} \propto M^{-2}$ (in Planck units), for the three lowest multipole numbers and all three responses. They also exhibit logarithmic running even for low multipole numbers in response to the scalar and vector fields. Such Planck scale suppression is similar to what was found in Ref. \cite{Kim:2020dif} and also for the AOS model, although the magnitude of the TLNs for the AOS model has comparatively weaker suppression, i.e., $\kappa_{\ell}^{s} \propto M^{-\frac{2}{3}}$. Specifically, the TLNs in response to a scalar field perturbation display a negative sign at leading order in the quantum parameter of the model, except for $\ell=0$, which has a vanishing TLN. This is similar to the AOS model, which predicts negative TLNs for the scalar field response. Furthermore, the leading order quantum gravitational corrections to the TLNs exhibit positive logarithmic running. For the vector field response, our results revealed that the leading order quantum gravitational corrections to the TLNs are negative for the first lowest multipole number $\ell=1$, while the next two multipole numbers, $\ell=2$ and $\ell=3$, have positive TLNs. We also found that all three TLNs exhibit positive logarithmic running at the leading order. This is in contrast to the AOS model, which predicts negative vector TLNs with no running for all three lowest multipole numbers. Finally, in the case of axial gravitational field perturbations, the leading order quantum gravitational corrections to the TLNs are all positive with no running for the three lowest multipole numbers, while the AOS model predicts negative tensor TLNs with no running in these cases.

In the case of the ZLMY-II loop quantum black hole, the TLNs are again nonzero for the three lowest multipole numbers and all three responses. They are also Planck scale suppressed, i.e., $\kappa_{\ell}^{s} \propto M^{-2}$, similar to the ZLMY-I model, and exhibit running behaviors at the leading order in the quantum parameter for the scalar and vector field responses. As a matter of fact, the scalar TLN is zero for $\ell=0$, similar to the ZLMY-I and AOS models, and it is negative with negative running for the next two lowest multipole numbers, $\ell=1$ and $\ell=2$, at the leading order. We also found that the leading order quantum gravitational corrections to the vector TLNs are negative for $\ell=1$, while they are positive for the next two lowest multipole numbers. Besides, they exhibit positive logarithmic running at the leading order, similar to the ZLMY-I model and in contrast to the AOS model with no running. Eventually, the leading order quantum gravitational corrections to the TLNs in response to the axial gravitational field perturbations are found to be positive without a running behavior, sharing similarities with ZLMY-I model and in contrast to the AOS model. 

For the ABV loop quantum black hole model, extracted TLNs from perturbative solutions demonstrate that they are nonzero and exhibit logarithmic running similar to both the ZLMY-I/II and opposed to the AOS model with no running. Although the magnitude of the TLNs is Planck scale suppressed, i.e., $\kappa_{\ell}^{s} \propto M^{-1}$, it is larger for the same black hole mass in comparison with ZLMY-I/II models. Regarding the sign and running behavior of the TLNs, in the case of scalar field response, we found that they are all negative for the three lowest multipole numbers, while they predict zero running for $\ell=0$ and negative running for $\ell=1$ and $\ell=2$ at the leading order in the quantum parameter, similar to the ZLMY-I/II models. On the other hand, the leading order quantum gravitational corrections to the TLNs, in response to the vector field perturbation, are all positive for the three lowest multipole numbers with vanishing running. This behavior also makes the ABV model distinct in comparison with ZLMY-I/II and AOS models. Finally, the tensor TLNs were found to be positive with zero running at leading order for all three multipole numbers considered in this study, similar to the ZLMY-I/II models.

In conclusion, the answer to the primary question  raised in this manuscript is in affirmative. Loop quantum black holes are generally deformed under the influence of the linear static external tidal field due to their nonzero TLNs. Our results also reveal that the quantum gravitational corrections to the TLNs are significantly stronger as the black hole mass reaches Planck scale mass, as is the case for loop quantum black holes, and are also in agreement with results found in Ref. \cite{Kim:2020dif}. These are intriguing results with potentially profound theoretical and phenomenological implications. Although tidal deformability is negligible for astrophysical black holes, it becomes stronger as the mass of the black hole reaches the Planck mass during the evaporation process. On one hand, such a tidal deformability might profoundly affect the final stage of the black hole evaporation by potentially modifying the emission spectrum of the black holes. On the other hand, if the external tidal field varies with time, the loop quantum black holes will acquire a time-dependent quadrupole moment, and they will tidally radiate gravitational waves \cite{Mashhoon-1973}. Due to Planck scale suppression of the TLNs, one expects that the rate of emission should be higher for Planck mass black holes. As the black holes evaporate and reach Planck mass, they also lose mass and energy through tidal radiation, possibly altering their lifetime.

Moreover, recent studies in classical GR and non-perturbative quantum gravity increase the possibility that black holes leave a remnant at the end of their evaporation \cite{Rovelli:2024sjl}. If that is the case, these Planck mass black hole remnants can account for dark matter in the cosmos. Since these particles only gravitationally interact, they have very small cross sections and cannot be directly detected with conventional experiments. However, it has been discussed in Ref. \cite{Christodoulou:2023hyt} that gravitationally interacting dark matter can be directly detected using the gravity-mediated quantum phase shift. Our results might modify this picture, implying that these particles, if they exist, have to be tidally deformed, and they must be extended (wavy) particles rather than point-like particles. Moreover, these particles are distinguishable from their internal composition. These features might affect the phenomenological properties of such dark matter candidates, leaving detectable observational effects.

Additionally, tidal deformability of loop quantum black holes indicates that the internal structure of quantum black holes is fundamentally distinct from that of their classical counterparts. Since the interior singularity of classical black holes is resolved in the presence of an effective repulsive force due to the discreteness of the spacetime geometry in the Planck regime, the nonzero TLNs also contain information about the dynamics of gravity at the Planck scale. Moreover, the tidal deformability of loop quantum black holes hints at the existence of a quantum hair, challenging the astrophysical version of the no-hair theorem \cite{Gurlebeck:2015xpa}. Crucially, this quantum hair is accessible to outside observers, and it can be quantified through TLNs. Such quantum hair may play an important role in solving the black hole information paradox by carrying information about the internal states of black holes beyond the horizon \cite{Essay}.

Finally, we should point out that although our results show that the tidal effects become stronger as the black hole mass reaches Planck scale mass, we are going beyond the validity of perturbative expansion for ZLMY-I/II and ABV models considered here and also the macroscopic limit for which the AOS model is valid. Hence, it would be useful to confirm the Planck scale suppression of TLNs for loop quantum black holes in a rigorous manner before moving forward to explore their phenomenological and theoretical implications. Furthermore, although we found that TLNs are nonzero for loop quantum Schwarzschild black holes, astrophysical black holes are described by the Kerr metric to include the spin of the black holes. Hence, it would be interesting to extend the current results to include the spin of the black holes using the rotating version of the considered models in this study built through the Newman-Janis algorithm \cite{Ban:2024qsa, Devi:2021ctm}. This will allow us to phenomenologically compare TLNs for loop quantum Kerr black holes with the classical Kerr black holes in GR and also understand the implications of the spin of black holes on the magnitude of TLNs for loop quantum black holes. These explorations will be carried out in future works. 

\section*{Acknowledgments}
 
This work is supported by NSF grant PHY-2409543. Authors thank Asier Alonso-Bardaji for useful discussions.

\end{document}